\documentclass[10pt,aps,prl,showpacs,preprint,twocolumn,superscriptaddress,floats,floatfix]{revtex4}
\usepackage{amssymb,amsmath}
\usepackage{graphics}
\usepackage{color,soul}
\usepackage{subfigure}
\usepackage{epsfig}
\usepackage{rotating}
\usepackage{float}
\usepackage{url}
\usepackage{hyperref}
\input symdef.inc.two
\begin{document}
\title{Universal Statistical Properties of Inertial-particle Trajectories in
Three-dimensional, Homogeneous, Isotropic, Fluid Turbulence}
\author{Akshay Bhatnagar}
\email{akshayphy@gmail.com}
\affiliation{Centre for Condensed Matter Theory, Department of Physics, Indian Institute of Science, Bangalore 560012, India.}
\author{Anupam Gupta}
\email{anupam@physics.iisc.ernet.in}
\affiliation{ University of Rome ``Tor Vergata'', Rome, Italy.}
\author{Dhrubaditya Mitra}
\email{dhruba.mitra@gmail.com}
\affiliation{NORDITA, Roslagstullsbacken 23, SE-10691 Stockholm, Sweden.}
\author{Prasad Perlekar}
\email{perlekar@tifrh.res.in}
\affiliation{TIFR Centre for Interdisciplinary Sciences, 21 Brundavan Colony, 
Narsingi, Hyderabad 500075, India}
\author{Rahul Pandit}
\email{rahul@physics.iisc.ernet.in}
\affiliation{Centre for Condensed Matter Theory, Department of Physics, Indian Institute 
of Science, Bangalore 560012, India.}

\begin{abstract}
We  uncover universal statistical properties
of the trajectories of heavy inertial particles in three-dimensional, statistically steady,
homogeneous, and isotropic turbulent flows by extensive direct numerical 
simulations.  We show that the probability distribution functions (PDFs)
$P(\phi)$, of the angle $\phi$ between the Eulerian velocity
${\bf u}$ and the particle velocity ${\bf v}$, at this point and time, shows a
power-law region in which $P(\phi) \sim \phi^{-\gamma}$, with a new
universal exponent $\gamma \simeq 4$. Furthermore, the  PDFs of the trajectory curvature $\kappa$ and modulus $\theta$
of the torsion $\vartheta$ have power-law tails that scale, respectively, as
$P(\kappa) \sim \kappa^{-h_\kappa}$, as $\kappa \to \infty$, and $P(\theta)
\sim \theta^{-h_\theta}$, as $\theta \to \infty$, with exponents $h_\kappa
\simeq 2.5$ and $h_\theta \simeq 3$ that are universal to the extent that they
do not depend on the Stokes number $\St$ (given our error bars).  
We also show that $\gamma$, $h_\kappa$ and $h_\theta$ can be obtained by
using simple stochastic models.
We characterize the complexity of heavy-particle trajectories by the number
$\NI(t,\St)$ of points (up until time $t$) at which $\vartheta$ changes sign.
We show that $\nI(\St) \equiv \lim_{t\to\infty} \frac{\NI(t,\St)}{t} \sim\St^{-\Delta}$,
with $\Delta \simeq 0.4$ a universal exponent.   
\end{abstract} 
\preprint{NORDITA-2014-114}
\maketitle

Inertial particles, advected by turbulent fluid flows, show rich dynamics that
are of great interest, not only because of potential applications in
geophysical~\cite{Csa73}, atmospheric~\cite{sha03,gra+wan13,fal+fou+ste02}, 
astrophysical~\cite{Arm10}, and industrial processes~\cite{eat+fes94,pos+abr02}, but also 
because they pose challenging questions of fundamental importance in the fluid
dynamics and nonequilibrium statistical mechanics of such flows. Experimental,
theoretical, and especially numerical investigations, which have been carried
out over the past few decades, have shown that neutrally buoyant tracers (or
Lagrangian particles) respond very differently to turbulent flows than do heavy,
inertial particles~\cite{toschirev,becoverview}; for instance, tracers get
distributed uniformly in space in a turbulent, incompressible flow, but, in the
same flow, heavy, inertial particles cluster~\cite{TOSCHI,Biferale}, especially
when the Stokes number $\St \simeq 1$, where $\St=\tau_s/\tau_\eta$,
with $\tau_s$ the particle-response or Stokes time and $\tau_\eta$ the
Kolmogorov time, at the dissipation length scale $\eta$. We study the
statistical properties of the geometries of heavy-inertial-particle
trajectories; such inertial-particle-trajectory statistics have not received
much attention hitherto in homogeneous, isotropic, three-dimensional (3D) fluid
turbulence. 

Our direct-numerical-simulation (DNS) studies of these statistical properties yield new
and universal scaling exponents that characterize heavy-particle trajectories. We calculate the
probability distribution functions (PDFs) of the angle $\phi$
between the Eulerian velocity ${\bf u}({\bf x},t)$, at the point ${\bf x}$ and
time $t$, and the velocity ${\bf v}$ of an inertial particle at this point and
time, PDFs of the curvature $\kappa$ and torsion
$\vartheta$ of inertial-particle trajectories, and several joint PDFs. In
particular, we find that the PDF $P(\phi)$ shows a power-law region
in which $P(\phi) \sim \phi^{-\gamma}$, with an exponent
$\gamma \simeq 4$, which has never been considered so far; the extent of this power-law 
regime decreases as $\St$
increases; we find good power-law fits if $0 < \St \lesssim 0.7$; in this range
$\gamma$ is universal, in as much as it does not depend on $\St$ and the fluid Reynolds number
 ${\rm Re}$ (given our error bars). The PDFs of $\kappa$ and $\theta=|\vartheta|$ show power-law
tails for large $\kappa$ and $\theta$, respectively, with power-law
exponents $h_\kappa$ and $h_\theta$ that are also universal. We calculate the number of 
points, per unit time, at
which the torsion $\vartheta$ changes sign along a particle trajectory; this
number $n_I(\St) \sim \St^{-\Delta}$, as $\St \to 0$, with $\Delta \simeq 0.4$
another universal exponent. We show how simple stochastic models can be used to 
obtain the exponents $\gamma$, $h_\kappa$, and $h_\theta$; however, the evaluation of
$\Delta$ requires the velocity field from the Navier-Stokes equation. 
  
\begin{table*}
\caption{Parameters for our runs ${\bf R1}$ and ${\bf R2}$ with
$N^3$ collection points, $\nu$ the coefficient of kinematic viscosity, $\delta
t$ the time step,  $N_p$ the number of particles, $k_{max}$ the largest
wave number in the simulation, $\eta$ and $\tau_\eta$ the dissipation
length and time scales, respectively, $\lambda$ the Taylor microscale,
$Re_\lambda$ the Taylor-microscale Reynolds number, $I_l$ the integral
length scale, and $T_{eddy}$ the large-eddy turnover time.}
\resizebox{0.9\linewidth}{!}{
\begin{tabular}{c c c c c c c c c c c c c}
\hline
Run & $N$ &	$\nu$	& $\delta t$ & $N_p$ & $Re_\lambda$ & $k_{max}\eta$ & $\epsilon$ &
$\eta$ & $\lambda$ & $I_l$ & $\tau_\eta$ & $T_{eddy}$ \\
\hline\hline
{\bf R1} & $256$ & $3.8\times 10^{-3}$ & $5\times10^{-4}$ & $40,000$ & $43$ & $1.56$ & $0.49$ &
$1.82\times10^{-2}$ & $0.16$ & $0.51$ & $8.76\times10^{-2}$ & $0.49$ \\

{\bf R2} & $512$ & $1.2\times 10^{-3}$ & $2\times10^{-4}$ & $100,000$ & $79$ & $1.21$ & $0.69$ &
$7.1\times10^{-3}$  & $0.08$ & $0.47$ & $4.18\times10^{-2}$ & $0.41$ \\
\hline
\end{tabular}}
\label{table:para}
\end{table*}

We perform a DNS~\cite{spectral} of the incompressible, three-dimensional (3D),
forced, Navier-Stokes equation
\begin{eqnarray}
\partial_t{\bf u} + ({\bf u}\cdot \nabla){\bf u} &=& \nu \nabla^2
{\bf u} - \nabla p +{\bf f}, \label{ns} \\
\nabla \cdot {\bf u} &=& 0, 
\label{eq:incom}
\end{eqnarray}
where ${\bf u}$, $p$, ${\bf f}$, and  $\nu$ are the velocity, pressure, external
force, and the kinematic viscosity, respectively.  Our simulation domain is a
cubical box with sides of length $2\pi$ and periodic boundary conditions in all
three directions. We use $N^3$ collocation points, a pseudospectral method
with a $2/3$ dealiasing rule~\cite{spectral}, a force that yields a constant
energy injection (see, e.g., Refs.~\cite{Lamorgese,Ganapati}), with an
energy-injection rate $P$, and a second-order Adams-Bashforth method for time
marching~\cite{Ganapati}. In several experiments (a) the radius of the particle
$\mathit{a} \ll \eta$, with $\eta$ the Kolmogorov dissipation scale of the advecting
fluid (i.e., the particle-scale Reynolds number is very small), (b) particle
interactions are negligible, (e.g., if the number density of particles is low),
(c) the particle density $\rho_p \gg \rho_f$, the fluid density, (d) typical
particle accelerations exceed considerably the acceleration because of gravity,
and (e) the particles do not affect the fluid velocity; if these conditions
hold, then the position ${\mathbf x}(t)$ and velocity ${\bf v}(t)$, at time
$t$, of a small, rigid, particle (henceforth, a heavy, inertial particle), in
an incompressible flow, evolve as follows~\cite{maxey83,gatignol83,bec2006acceleration}:

\begin{figure*}
\centering
\resizebox{\linewidth}{!}{
\includegraphics[width=0.3\linewidth]{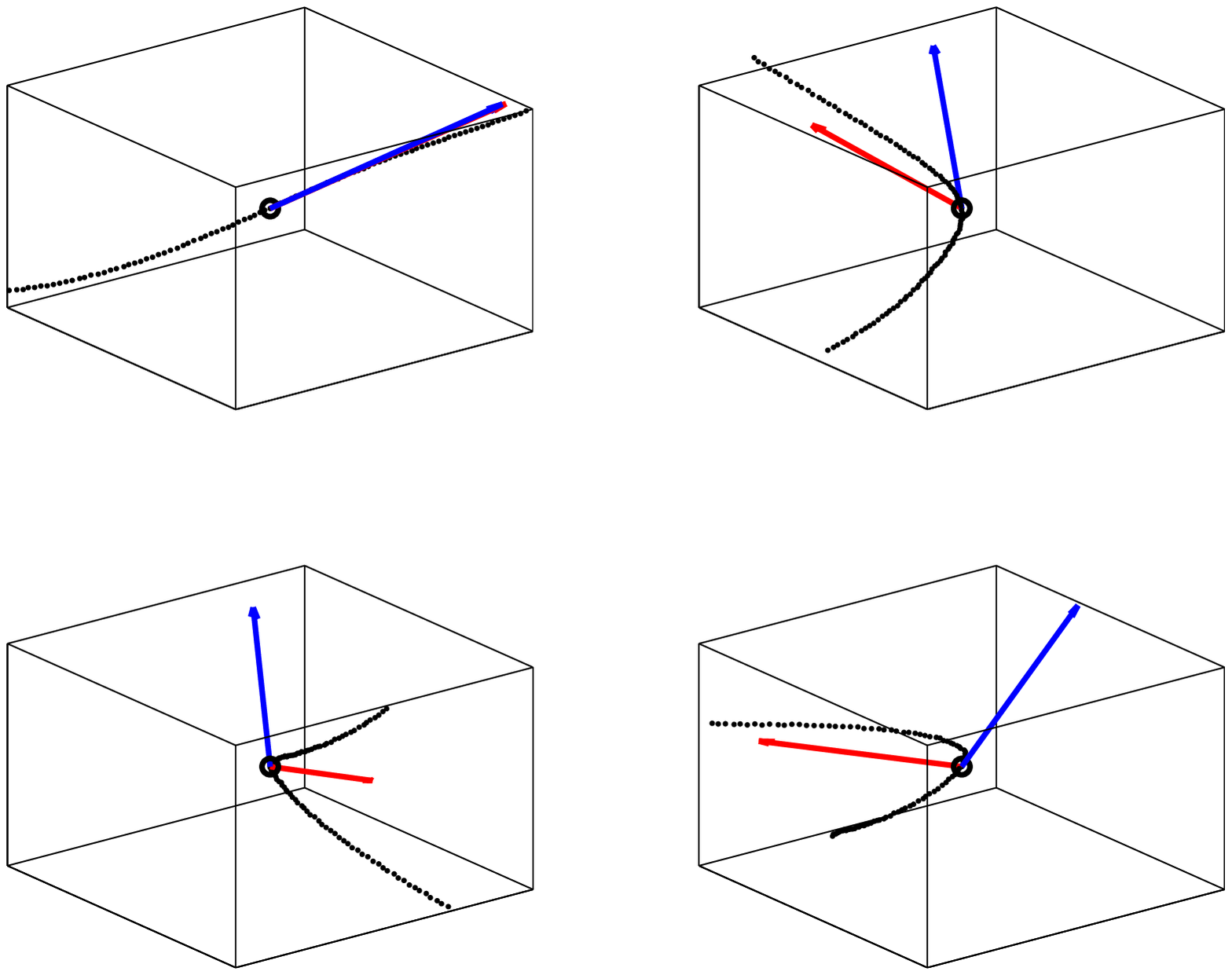}
\put(-80,60){\bf (a)}
\includegraphics[width=0.3\linewidth]{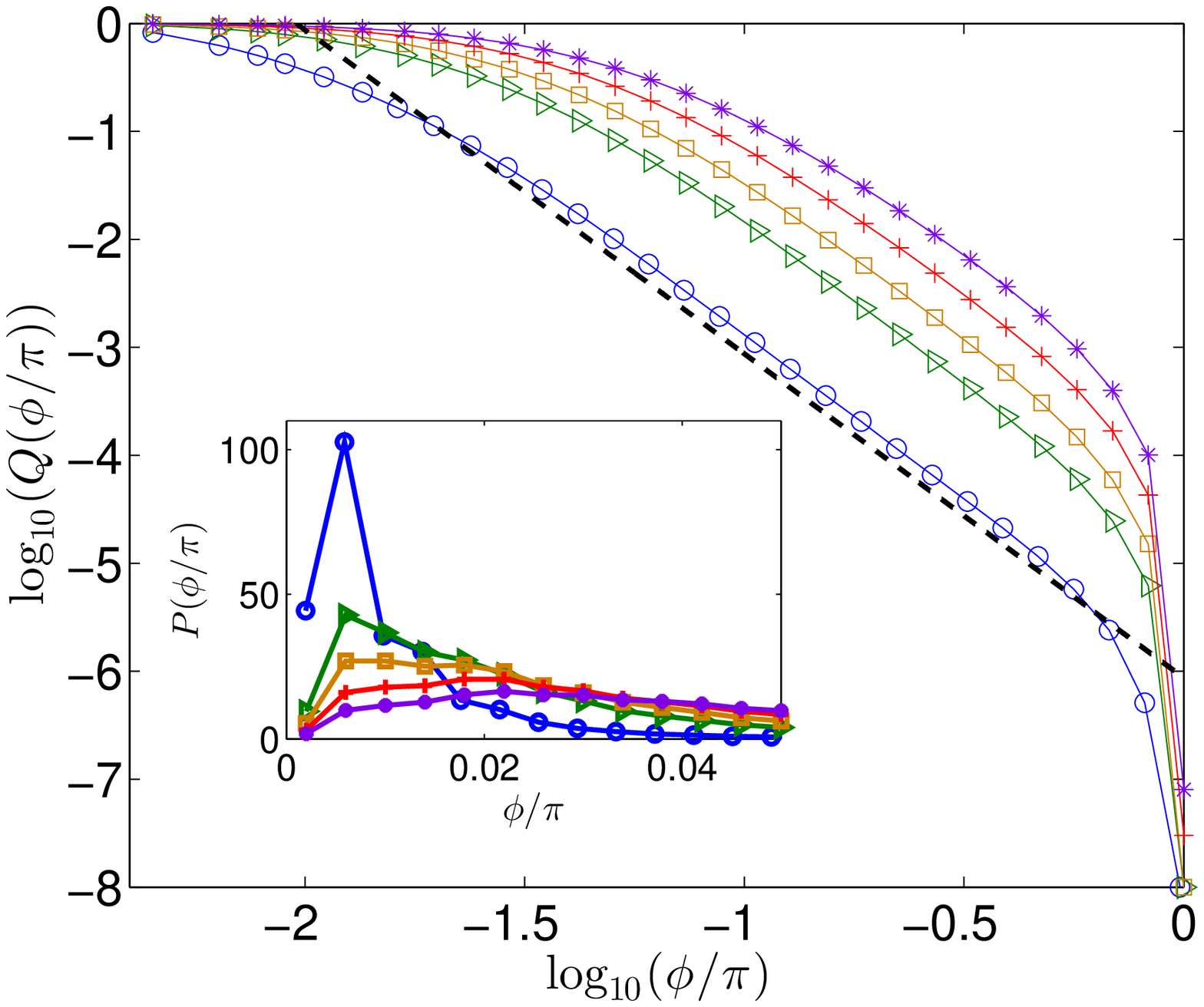}
\put(-25,100){\bf (b)}
\includegraphics[width=0.3\linewidth,height=0.26\linewidth]{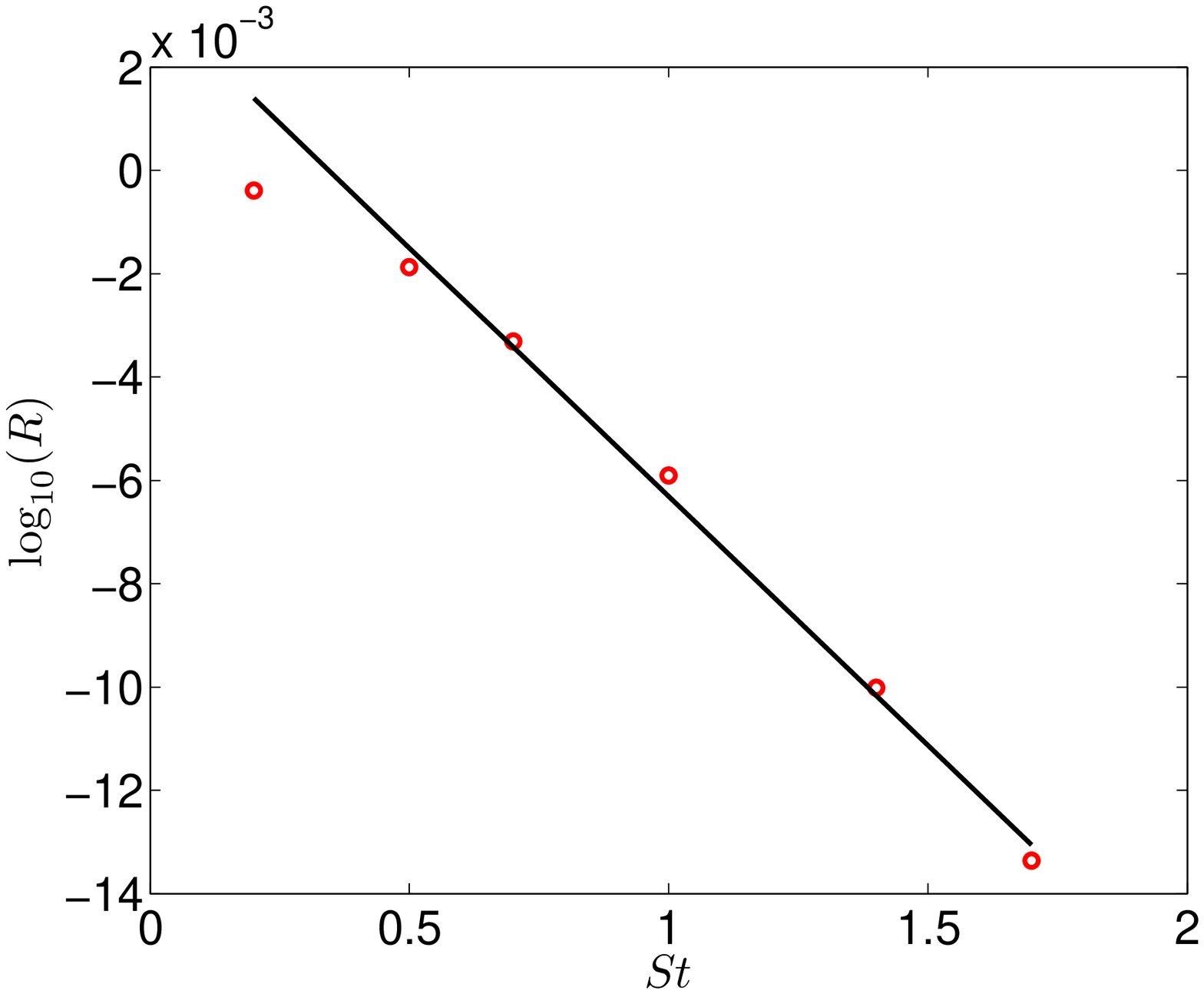}
\put(-25,100){\bf (c)}}
\caption{(Color online) (a) Plots of the unit vectors along the directions of the particle and fluid
velocities (in blue and red, respectively) at the position of the particle, 
and the trajectory of the particle in the
neighbourhood of the particle position (black dots), at four different time instants for
$\St=1.0$. 
(b) Cumulative PDFs of the angle $\phi$ between $\mathbf{u}$ and $\mathbf{v}$ 
($Q(\alpha)\equiv P(\phi\geq \alpha)$), for $\St = 0.2$ (blue circles), $\St = 0.5$ (green
triangles), $\St = 0.7$ (brown squares), $\St = 1.0$ (red pluses), and $\St = 1.4$ (purple
stars), 
obtained by using the rank-order method ~\cite{dmitra}; the slope of the black dashed line
is $-3$; the 
inset shows plots of PDF $P(\phi)$, 
(c) coefficient of correlation $R$ between $u_x$ and $v_x$  as a function of $\St$ (red
circles) obtained from DNS data, black line shows the exponential fit.}
\label{fig:anguv}
\end{figure*}

\begin{eqnarray}
\frac{d}{dt}{\bf v}(t) &=& - \frac{{\bf v}(t) - {\bf u}({\bf x},t)}{\tau_s};
\label{eq:part1}
\end{eqnarray}
and
\begin{eqnarray}
\frac{d}{dt}{\mathbf x}(t) &=& {\bf v}(t);
\label{eq:part2}
\end{eqnarray}   
here ${\bf u}({\bf x},t)$ denotes the Eulerian velocity field at position ${\bf
x}$ and $\tau_s = (2 \mathit{a}^2)/(9\nu \rho_{\rm f})$. 
To obtain the statistical properties of the particle paths, we
follow the trajectories of $N_p$ particles in our simulation, use trilinear
interpolation~\cite{num_rec} to calculate the components of the velocity and
the velocity-gradient tensor at the positions of the particles. Table
\ref{table:para} lists the parameters we use.  We solve for the trajectories of
$N_p$ inertial particles, for each of which we solve
Eqs.(\ref{eq:part1}) and (\ref{eq:part2}) with an Euler scheme, which is adequate because,
in time $\delta t$, a particle crosses at most one-tenth of the grid spacing.  

At the position of a particle, the particle and fluid velocities are different
because of the Stokes drag. We expect this difference to increase with $\St$.
In Fig.~\ref{fig:anguv} (a) we show plots of the unit vectors, along the
directions of the particle and fluid velocities (in red and green,
respectively), at the position of the particle, and the trajectory of the
particle in the neighborhood of the particle position (blue dots), at four
different time instants (for a video see Ref.~\cite{supp}). {To quantify the
statistics of the angle between these unit vectors, we show in the inset of Fig.~\ref{fig:anguv} (b) 
plots of the PDFs $P(\phi)$ of the angle $\phi$ between $\mathbf{\hat{v}}$ and 
$\mathbf{\hat{u}}$, for different values of $\St$. For small $\St$, 
$P(\phi)$ shows a peak near $\phi \simeq 0$; this peak broadens  
when we increase $\St$. Log-log plots of the cumulative PDFs $Q(\phi)$ 
[Fig.~\ref{fig:anguv} (b)] reveal that, especially 
for small values of $\St$, there is a remarkable and distinct power-law regime 
in which $Q(\phi) \sim
\phi^{-\gamma+1}$,} with a scaling exponent $\gamma \simeq 4$, i.e., the
PDF $P(\phi) \sim \phi^{-4}$.  Although $\gamma$ is insensitive to
the value of $\St$ (given our error bars), the extent of the scaling regime
decreases as we increase $\St$. 

A particle trajectory is a 3D curve that we characterize by its tangent
${\mathbf t}$, normal ${\mathbf n}$, and binormal ${\mathbf b}$, which
are~\cite{diff_geom,diff_geom2,wbraun}
\begin{equation}
{\mathbf t}=\frac{d{\mathbf r}}{d s}; \; {\mathbf n} = \frac{1}{\kappa}\frac{d{\mathbf
t}}{d s}; \; 
{\mathbf b} = {\mathbf t}\times {\mathbf n};
\label{eq:tnb}
\end{equation}
these evolve according to the Frenet-Serret formulas as:
\begin{equation}
\frac{d {\mathbf t}}{ds} = \kappa {\mathbf n}; \;
\frac{d {\mathbf n}}{ds} = \vartheta {\mathbf b} - \kappa {\mathbf t}; \;
\frac{d {\mathbf b}}{ds} = -\vartheta {\mathbf n;}
\label{eq:tnb2}
\end{equation}
here ${\mathbf r}$ and $s$ indicate the particle position and arc length along the 
particle trajectory, respectively. The curvature 
$\kappa$ and the torsion $\vartheta$ of a particle trajectory are (dots indicate
time derivatives)
\begin{equation}
\kappa = \frac{|{\bf v} \times {\bf \dot{v}}|}{|{\bf v}|^3} = a_n/v^2; \;
\vartheta = \frac{{\bf v} \cdot ({\bf \dot{v}} \times {\bf \ddot{v}})}{({\bf v}\cdot{\bf v})^3
\kappa^2}.
\label{eq:kaptor}
\end{equation}
Here $a_n$ is the normal component of the acceleration $\mathbf a$, $v = |\bf v|$, and
$\theta = |\vartheta|$. {From Eqs.(\ref{eq:part1}) and (\ref{eq:tnb2}) it follows
that} ${\mathbf u}$, ${\mathbf v}$, and ${\mathbf a}$
can be expressed as
${\mathbf u} = u\cos{\phi} {\mathbf t}+u\sin{\phi} {\mathbf n}$, 
${\mathbf v} = v{\mathbf t}$, and ${\mathbf a} = a_t {\mathbf t}+a_n {\mathbf n}$,
whence we obtain
\begin{equation}
\sin{\phi} = \frac{a_n\tau_s}{u} = \frac{\kappa v^2 \tau_s}{u}.
\label{eq:kap_phi}
\end{equation}

\begin{figure*}
\centering 
\resizebox{\linewidth}{!}{
\includegraphics[width=0.33\linewidth]{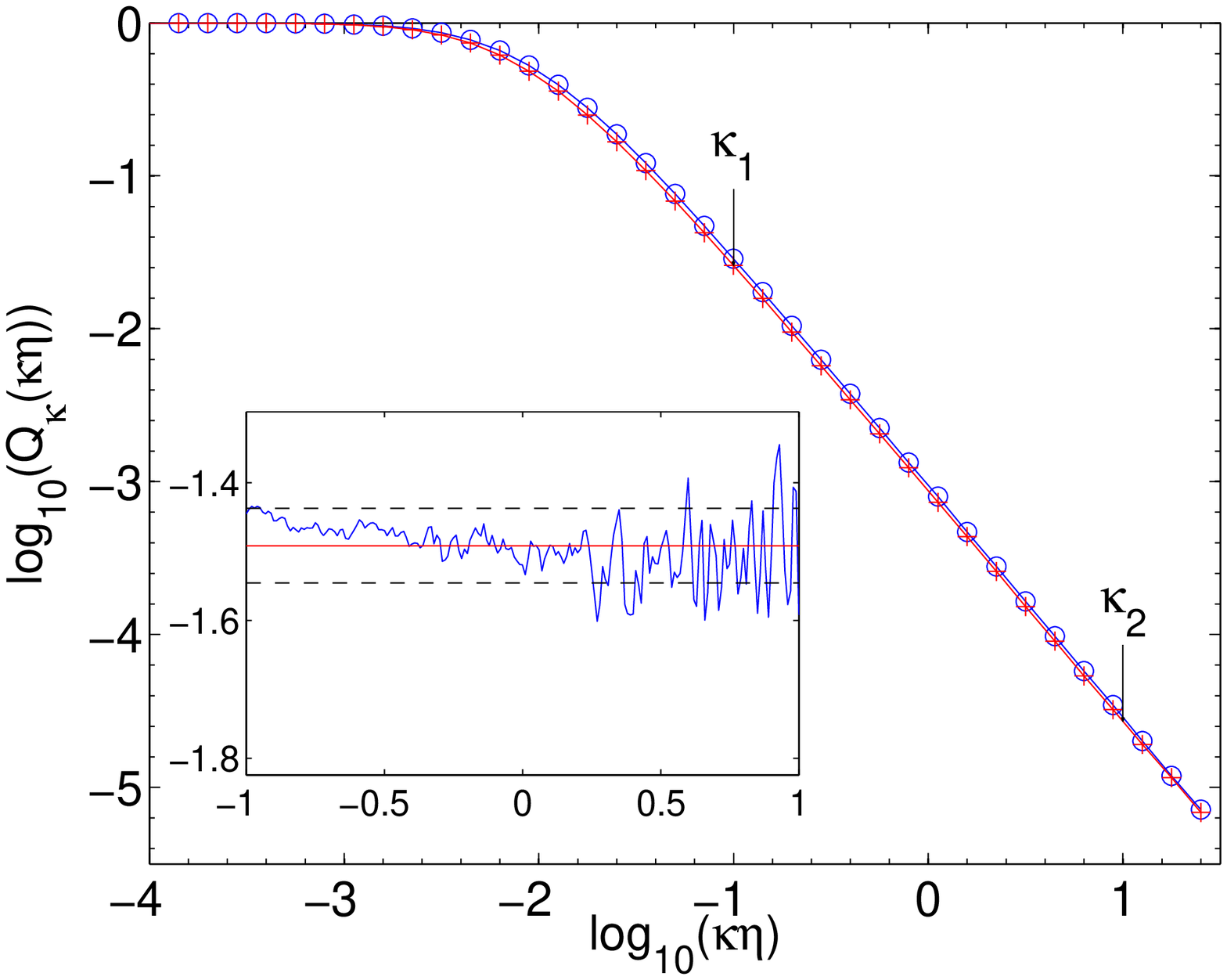}
\put(-25,110){\bf (a)}
\includegraphics[width=0.33\linewidth]{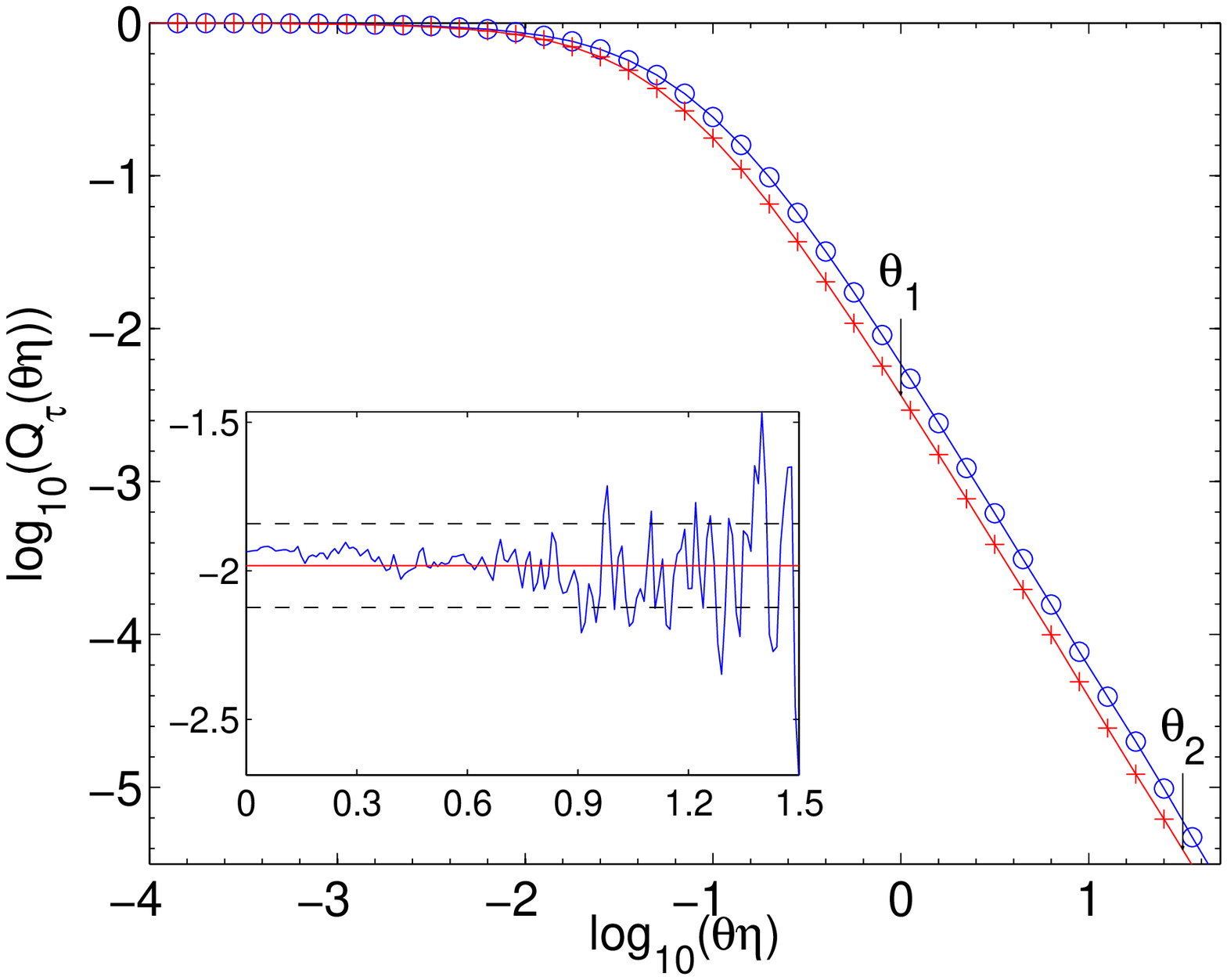}
\put(-25,110){\bf (b)}
\includegraphics[width=0.33\linewidth]{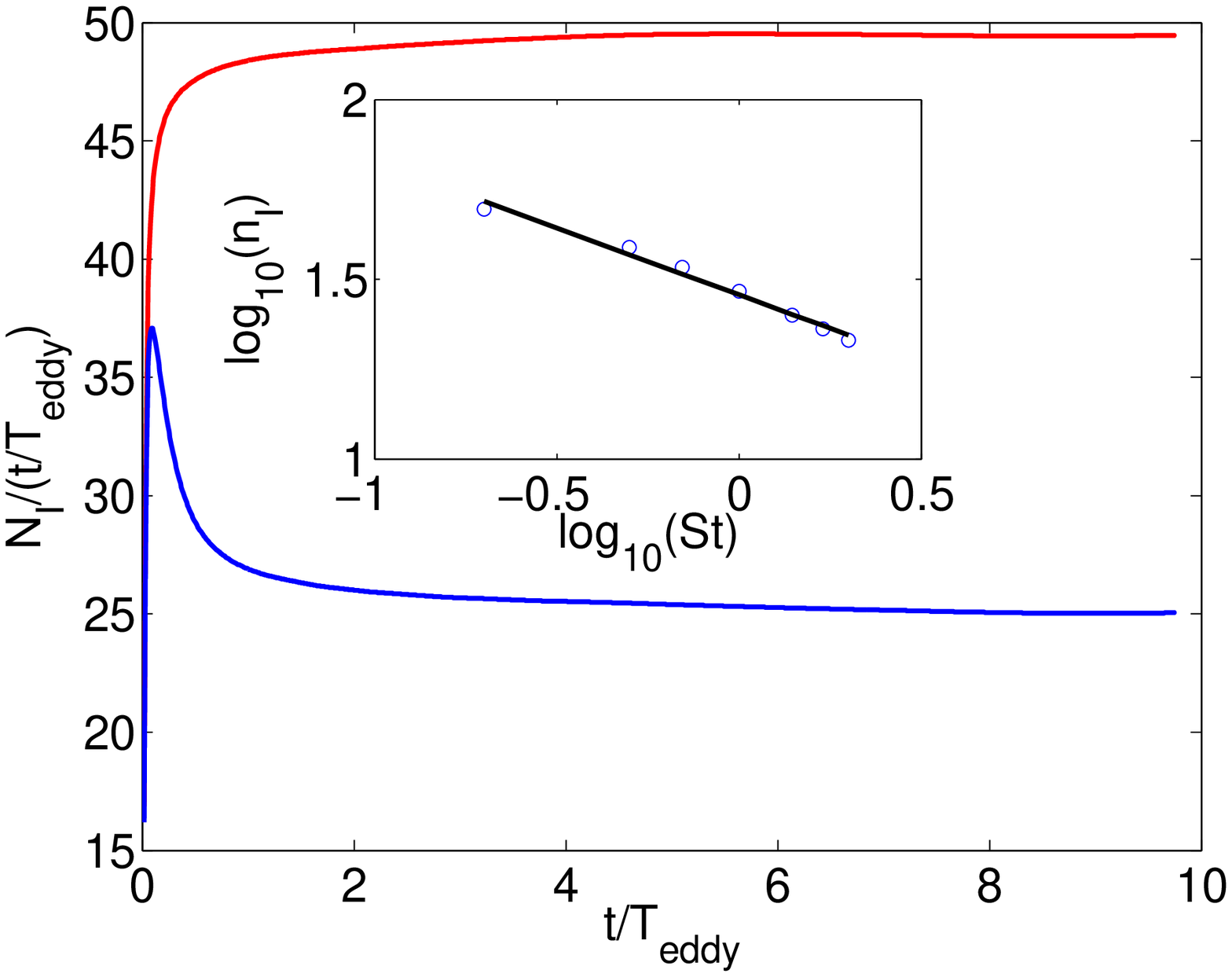}
\put(-25,110){\bf (c)}}
\caption{(Color online) Cumulative PDFs of, (a) the curvature $\kappa$ and (b) 
the magnitude of the torsion $\theta$ of the 
trajectories of heavy inertial particles, for $St = 0.2$ (in blue) and $1.0$ (in red), 
obtained using rank order method. Inset: the values of the local slope of the tail, 
for $St = 1.0$. (c) Number of inflection points per unit time (see Eq. (7)) 
as a function of dimensionless time
$t/T_{eddy}$, for $St=0.2$, (red curve), and $St=1.4$, (blue curve); the inset shows the plot
of the number of inflection points per unit time $n_I$, as a function of $\St$, obtained 
from run {\bf R2}}
\label{fig:cpdfkappa}
\end{figure*}

We find, in agreement with Ref.~\cite{bec2006acceleration}, that the PDFs of the normal 
component $a_n$ the tangential component $a_t$ and 
$a=|\mathbf{a}|$ exhibit tails~\cite{supp}, which can be fit to
exponential forms with decay rates $\alpha_n$, $\alpha_t$, and $\alpha$,
respectively; these decay rates decrease as $\St$ increases (Table
~\ref{table:exponents}). By contrast, the PDFs $P(\kappa)$ and $P(\theta)$ have
power-law tails that scale as $P(\kappa) \sim \kappa^{-h_\kappa}$, as $\kappa
\to \infty$, and $P(\theta) \sim \theta^{-h_\theta}$, as $\theta \to \infty$,
with exponents $h_\kappa \simeq 2.5$ and $h_\theta \simeq 3$ that do not
depend on $\St$ (given our error bars) \footnote{$P(\kappa) \sim
\kappa^{h_l}$, as $\kappa \to 0$}. We obtain these exponents accurately from the
cumulative PDFs $Q_\kappa(\kappa\eta)$ and $Q_\tau(\theta\eta)$, which we
obtain by using a rank-order method~\cite{dmitra} to overcome binning errors
and which we show in the log-log plots of Figs.~\ref{fig:cpdfkappa} (a) and
(b), respectively, for representative values of $\St$. The slopes of the
straight-line parts (blue lines) in these plots yield $h_\kappa+1$ and
$h_\theta+1$; we list $h_\kappa$ and $h_\theta$ in
Table~\ref{table:exponents}; to obtain the error bars on these exponents we
carry out a local-slope analysis~\cite{Perlekar} for the power-law regimes in  these cumulative
PDFs (see the insets of Figs.~\ref{fig:cpdfkappa} (a) and (b)). 

We use the torsion $\vartheta$ to characterize the complexity of a particle
track by computing the number, $N_I(t,\St)$, of points at which 
$\vartheta$ changes sign up until time $t$. We propose that, for a given
value of $\St$,
\begin{equation}
n_I(\St) \equiv \lim_{t \to \infty} \frac{N_I(t,\St)}{t}
\label{eq:inf}
\end{equation}
exists and is a natural measure of its complexity. 
In Fig.~\ref{fig:cpdfkappa} (c), we plot $N_I/(t/T_{eddy})$
versus the dimensionless time $t/T_{eddy}$, for
two representative values of $\St$. From such plots we obtain
$n_I(\St)$ (see Eq.~(\ref{eq:inf})), which we depict as a function of $\St$
in the inset of Fig.~\ref{fig:cpdfkappa} (c), and whence we find
\begin{equation}  
n_I(\St) \sim \St^{-\Delta},
\end{equation}  
where $\Delta \simeq 0.4$. This indicates that, as $\St \to 0$,
particle trajectories become more and more contorted in all three
spatial dimensions (cf. Ref.~\cite{geometry2d} for the analog of
this result for 2D fluid turbulence).

The analogs of the exponents  $h_\kappa$
and $h_\theta$ have been obtained in studies with Lagrangian tracers in
DNSs~\cite{wbraun,scagliarini} and experiments~\cite{curvatureprl}. The
exponents obtained in both these studies, for tracers, are within error bars of
those that we obtain here for heavy inertial particles (see
Table~\ref{table:exponents}).  In Refs.~\cite{scagliarini,curvatureprl} it has
been suggested that the value of $h_\kappa$ can be obtained by
noting that large values of $\kappa$ are associated with small values of
$v$; furthermore, $h_\kappa = 2.5$ can be obtained analytically
by assuming that the joint PDF $\mathcal{P}(a_n,v)$ factors into the products
of the PDFs $P(a_n)$ and $P(v)$. A similar argument~\cite{scagliarini} yields
$h_\theta = 3$. These arguments can be extended to the case of heavy inertial
particles and used, therefore, to understand the proximity of the values of
$h_\kappa$ and $h_\theta$  (see Table~\ref{table:exponents}) to those for
Lagrangian tracers. 
In Fig.~\ref{fig:jpdf}, we plot joint PDFs of $\kappa$ and $v$, and $\kappa$ and $a_n$
 for two representative values of $\St$ ($\St =
0.2$ left column, $\St = 1.4$ right column). Clearly, large values of $\kappa$ are
correlated with small values of $v$ but not with large values of $a_n$; i.e.,
high-curvature parts of particle trajectories are associated with regimes of a track
where the particle velocity reverses. Furthermore, we show in ~\cite{supp} that the
assumption $\mathcal P(a_n,v) \equiv P_{an}(a_n) P_v(v)$ made in ~\cite{scagliarini} does
not hold very well.   

{To understand the universal power laws mentioned above,
we use a simple stochastic model for the Eulerian velocity field~\cite{supp}, and integrate
Eqs.(\ref{eq:part1}) and (\ref{eq:part2}) to find particle trajectories. We find that such a simple
model, in which the Eulerian velocity field is given by Eqs.(4)-(6) in~\cite{supp}, 
reproduces 
the exponents $\gamma$, $h_\kappa$, and $h_\theta$ accurately~\cite{supp} \textit{but not} $\Delta$. 

We also show
numerically that, if we consider the components of fluid and particle velocities 
$u_i$ and $v_j$, where $i,j \in (x,y,z)$, to be correlated Gaussian
random variates with mean zero, such that the coefficient of correlation between them $\rho(u_i,v_j)
\equiv \langle u_i v_j \rangle/\sqrt{\langle u_i^2 \rangle \langle v_j^2 \rangle}$ is a
function of $\St$, such that, $\rho(u_i,v_j) = R(\St) \delta_{ij}$, then we get the same values of
the exponents, $\gamma$, $h_\kappa$, and $h_\theta$ 
~\cite{supp} as in Table~\ref{table:exponents}. The plot of $\log_{10}(R)$ versus
$\St$ in Fig.~\ref{fig:anguv} (c), shows that $R$ decays
exponentially with increasing $\St$ in our DNS of Eqs.~(\ref{ns})-(\ref{eq:part1}).

We hope that our results will stimulate new experimental
studies~\cite{ewsaw} of the geometries of inertial-particle trajectories in turbulent
flows. 
Our results for $P(\phi)$, $P(\kappa)$, $P(\theta)$, and $n_I(\St)$ 
can be used to constrain models for the statistical properties of inertial particles 
in turbulent flows~\cite{model1,model2}. In particular, we show that simple stochastic 
models can yield
$\gamma$, $h_\kappa$, and $h_\theta$ but not the exponent $\Delta$
Ref.~\cite{supp}.

The exponent $\Delta$ has not been
introduced in 3D fluid turbulence so far.  Our results imply that $\nI(\St)$
has a power-law divergence, as $\St \to 0$.  This is suppressed eventually, in
any finite-resolution DNS, which can only achieve a finite value of
$Re_{\lambda}$. This is the analog of the finite-size suppression of
divergences, in thermodynamic functions, at an equilibrium critical
point~\cite{fssprivman}. Furthermore, the limit $\St \to 0$ is singular, so it
is not clear {\it a priori} that it should yield the same results, for the
properties we study, as those in the Lagrangian case $\St =0$.

\begin{figure}
\includegraphics[width=0.49\linewidth]{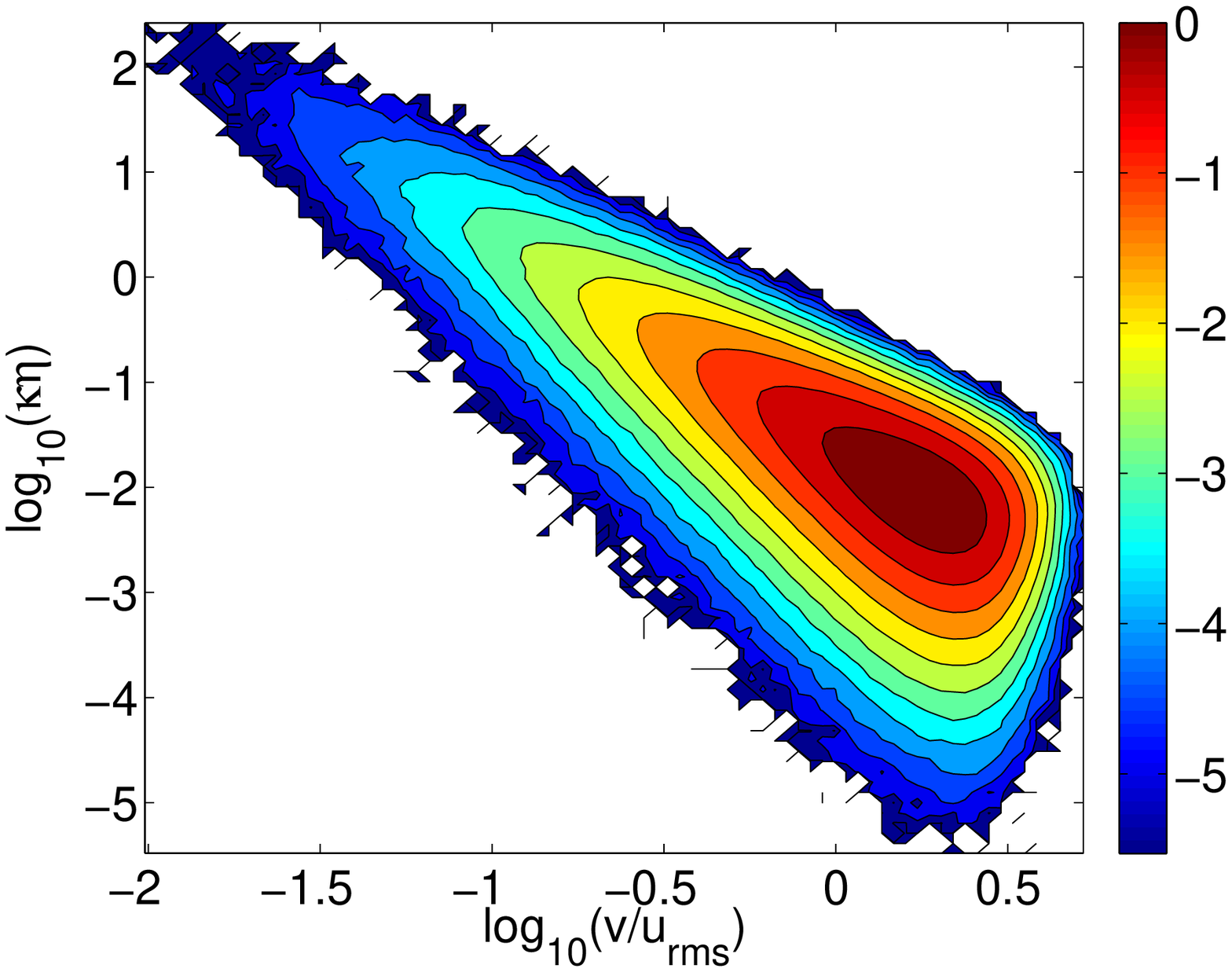}
\put(-75,95){\bf $St = 0.2$}
\includegraphics[width=0.49\linewidth]{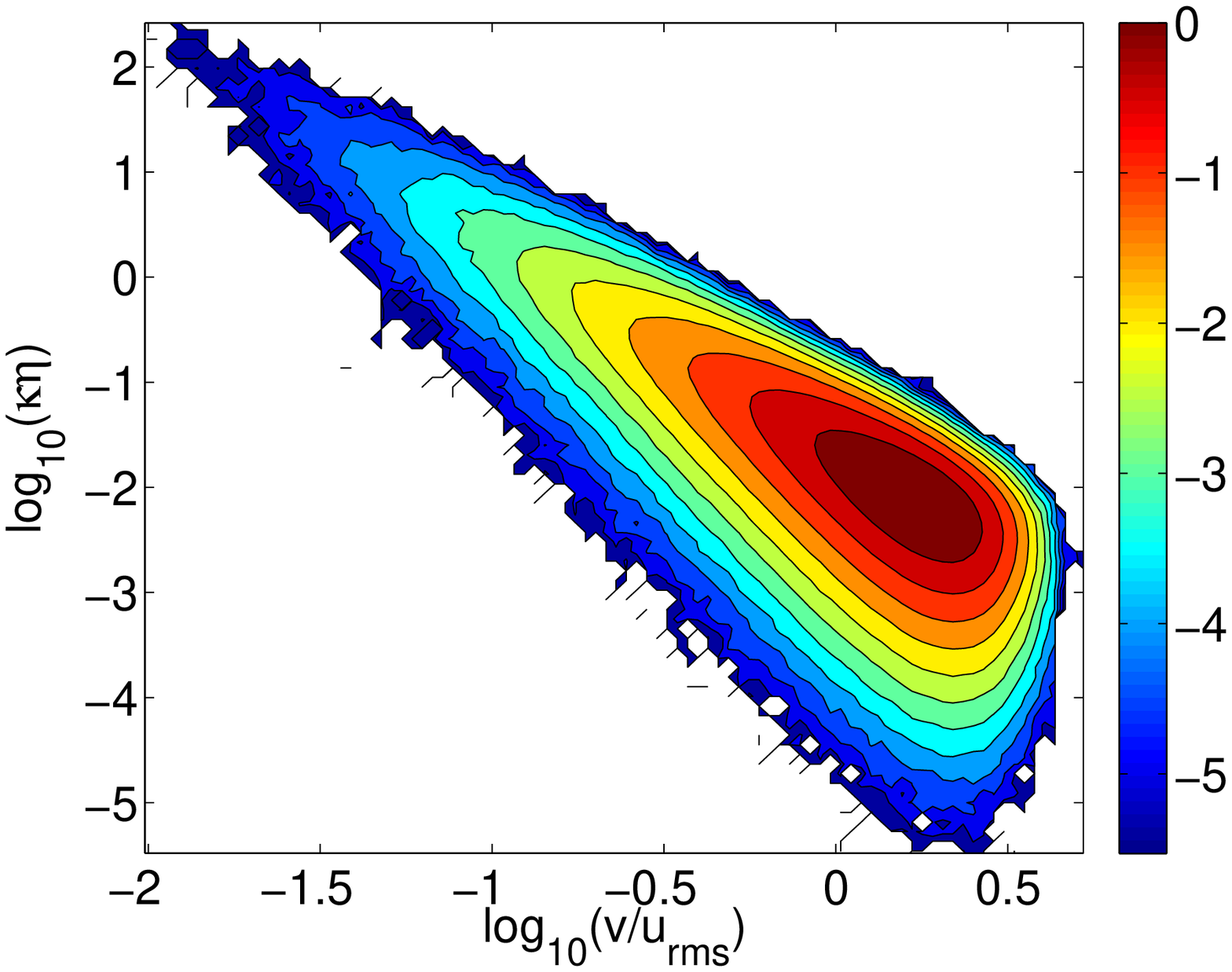}
\put(-75,95){\bf $St = 1.4$}\\
\includegraphics[width=0.49\linewidth]{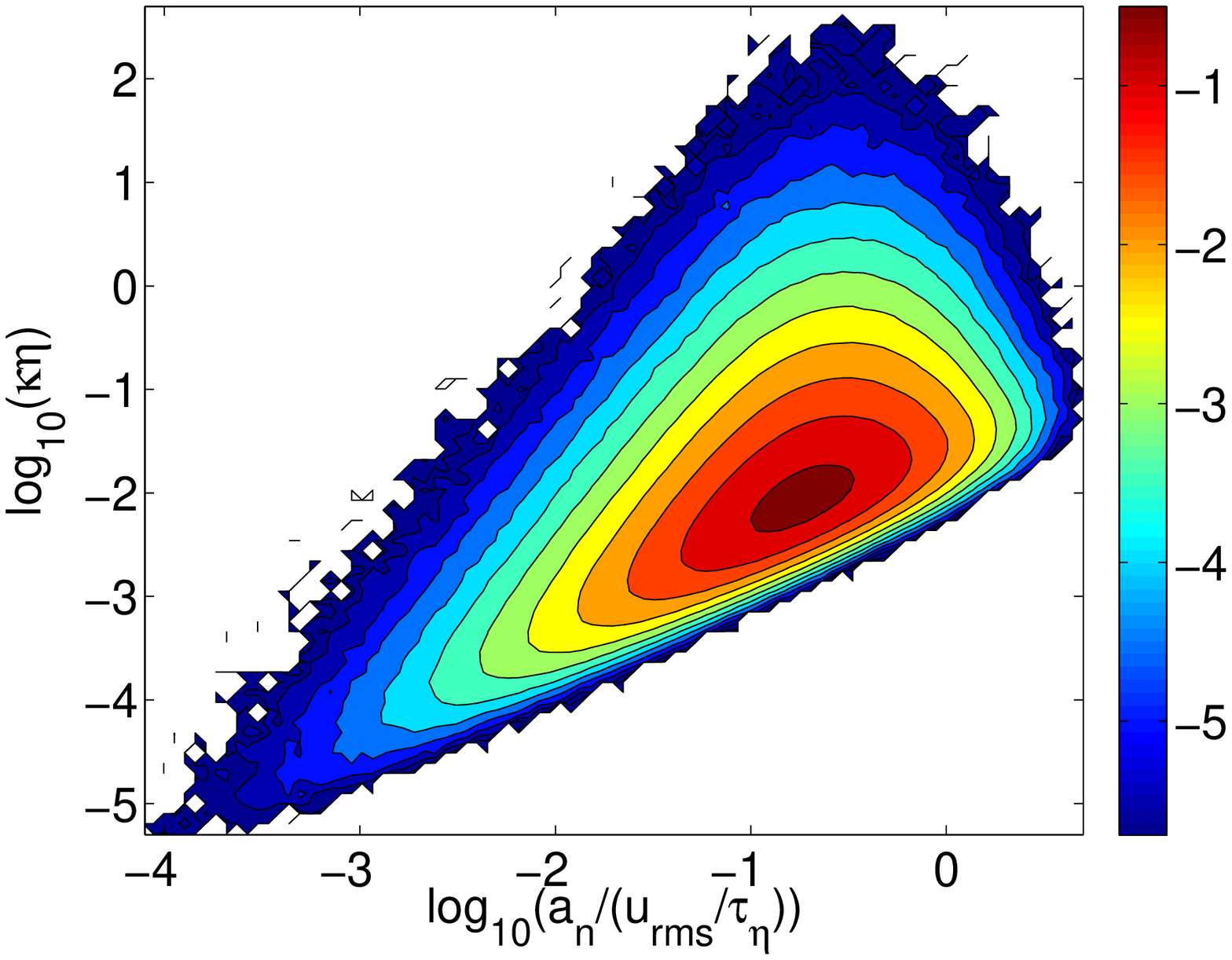}
\includegraphics[width=0.49\linewidth]{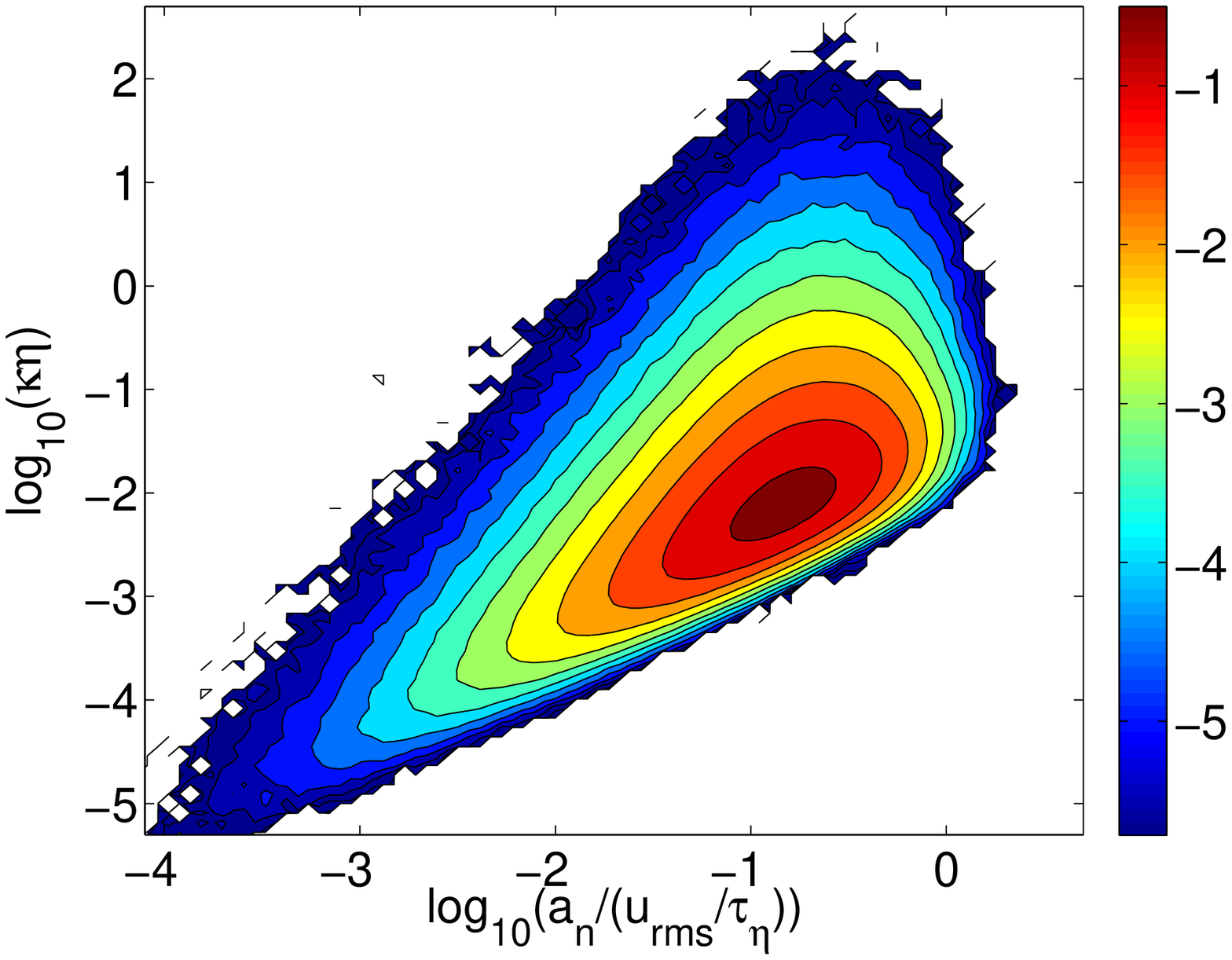}
\caption{(Color online) Filled contour plots of the joint PDFs of the curvature and velocity
of the particle (top row) and curvature and normal component of the particle
acceleration (bottom row).} 
\label{fig:jpdf}
\end{figure}

\begin{table}
\caption{Decay constants $\alpha$, $\alpha_t$ and $\alpha_n$, from the exponential fit to the
PDFs of the acceleration, tangential component of acceleration and normal component of the
acceleration, respectively, and the exponents $h_\kappa$ and $h_\theta$ from the power-law
fits
to the tails of the cumulative PDFs of the curvature $\kappa$ and torsion $\theta$,
respectively, for
the different values of $\St$.}
\begin{tabular}{c c c c c c}
\hline
$\St$ & $\alpha$ & $\alpha_t$ & $\alpha_n$ & $h_\kappa$ & $h_\theta$ \\
\hline\hline
$0.2$ & $0.31\pm0.08$ &  $0.19\pm0.09$ &  $0.30\pm0.07$ & $2.4\pm0.2$		& $3.0\pm0.3$ \\
$0.5$ & $0.21\pm0.08$ &  $0.13\pm0.08$ &  $0.20\pm0.09$ &	$2.6\pm0.3$		&	$3.1\pm0.3$ \\
$0.7$ & $0.18\pm0.06$ &  $0.11\pm0.09$ &  $0.17\pm0.08$ &	$2.5\pm0.2$		&	$3.0\pm0.2$ \\
$1.0$ & $0.15\pm0.04$ &  $0.10\pm0.07$ &  $0.14\pm0.06$ &	$2.48\pm0.09$	&	$2.9\pm0.2$ \\
$1.4$ & $0.13\pm0.05$ &  $0.08\pm0.06$ &  $0.11\pm0.06$ &	$2.4\pm0.1$		&	$3.1\pm0.2$ \\
\hline
\end{tabular}
\label{table:exponents}
\end{table}

We thank J. Bec, A. Brandenburg, B. Mehlig, E.W. Saw, and D. Vincenzi
for discussions, and particularly A. Niemi, whose study of the intrinsic
geometrical properties of polymers~\cite{poly11} inspired our work on particle
trajectories, and S. S. Ray for an introduction to the first of the stochastic models we
use in Ref~\cite{supp}.  This work has been supported  in part by the European Research
Council under the AstroDyn Research Project No.\ 227952 (DM), Swedish Research
Council under grant 2011-542 (DM), NORDITA visiting PhD students program (AG),
and CSIR, UGC, and DST (India) (AB, AG and RP).  We thank SERC (IISc) for providing
computational resources. AG, PP, and RP thank NORDITA for hospitality under their
Particles in Turbulence program; DM
thanks the Indian Institute of Science for hospitality.


\clearpage

{\large\bf{Supplemental Material}}
\vspace{0.5cm}

This Supplemental Material contains some probability distribution functions (PDFs) and
joint PDFs that augment the figures given in the main part of this paper.
We give semilogarithmic plots of the PDFs  $P_\mathrm{a}$, $P_\mathrm{at}$, 
and $P_\mathrm{an}$ of the
acceleration (Fig.~\ref{fig:acc} (a)), its tangential component
(Fig.~\ref{fig:acc} (b)), and its normal component (Fig.~\ref{fig:acc}
(c)), respectively. The right tails of these PDFs can be fit to
exponential forms; in particular,
\begin{eqnarray}
P_\mathrm{a}(a/(u_{rms}/\tau_\eta)) &\sim& \exp\left(-\frac{a/(u_{rms}/\tau_\eta)}{\alpha}\right),\\
P_\mathrm{at}(a_t/(u_{rms}/\tau_\eta)) &\sim& \exp\left(-\frac{a_t/(u_{rms}/\tau_\eta)}{\alpha_t}\right),\\
P_\mathrm{an}(a_n/(u_{rms}/\tau_\eta)) &\sim& \exp\left(-\frac{a_n/(u_{rms}/\tau_\eta)}{\alpha_n}\right),
\label{eq:acc}
\end{eqnarray}
with decay rates $\alpha$, $\alpha_t$, and $\alpha_n$ whose values we
list, for different values of the Stokes number $\St$, in TABLE ~\ref{table:exponents}.  All
these decay rates decrease as $\St$ increases.

\begin{figure*}
\includegraphics[width=0.3\linewidth]{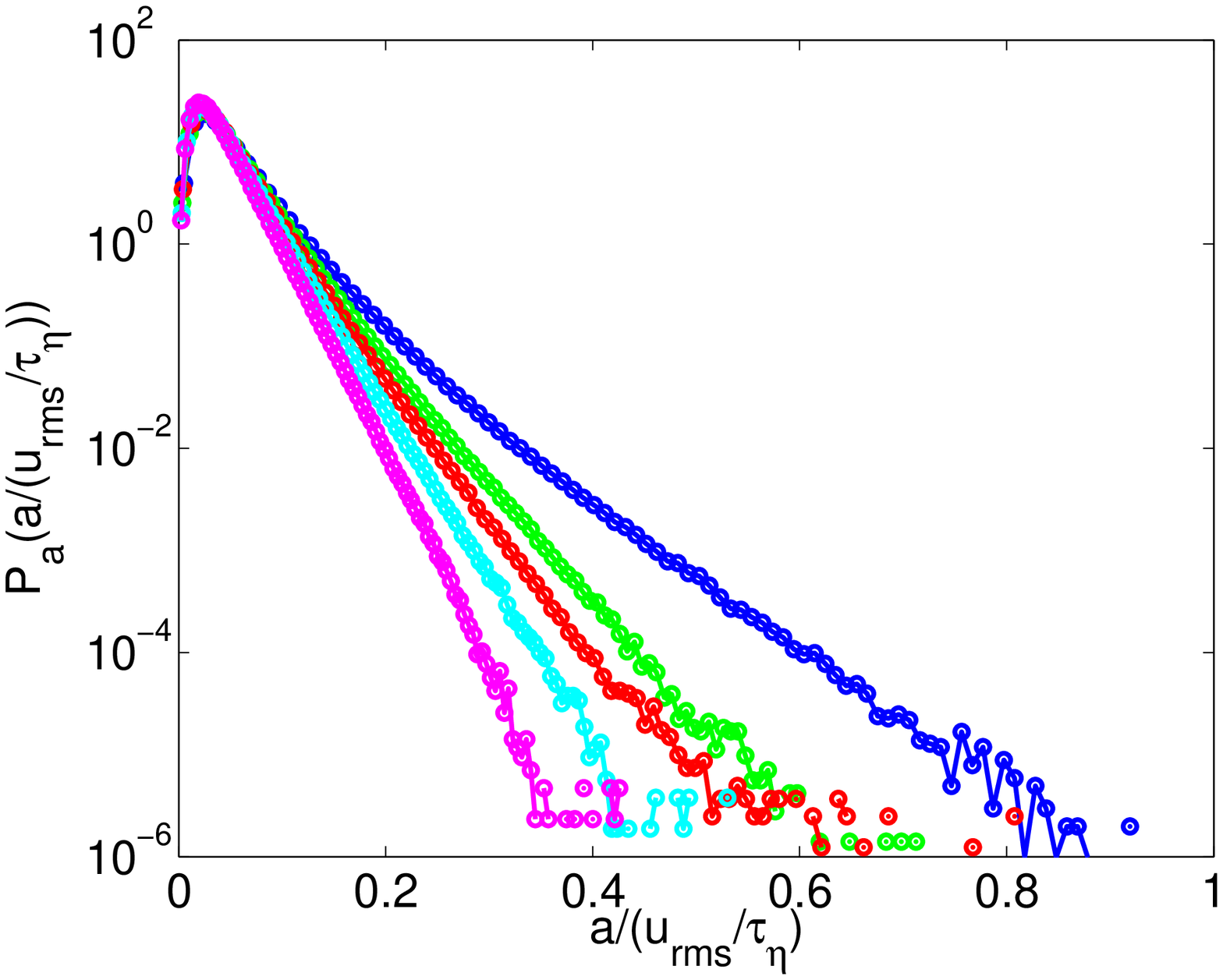}
\put(-25,100){\bf (a)}
\includegraphics[width=0.3\linewidth]{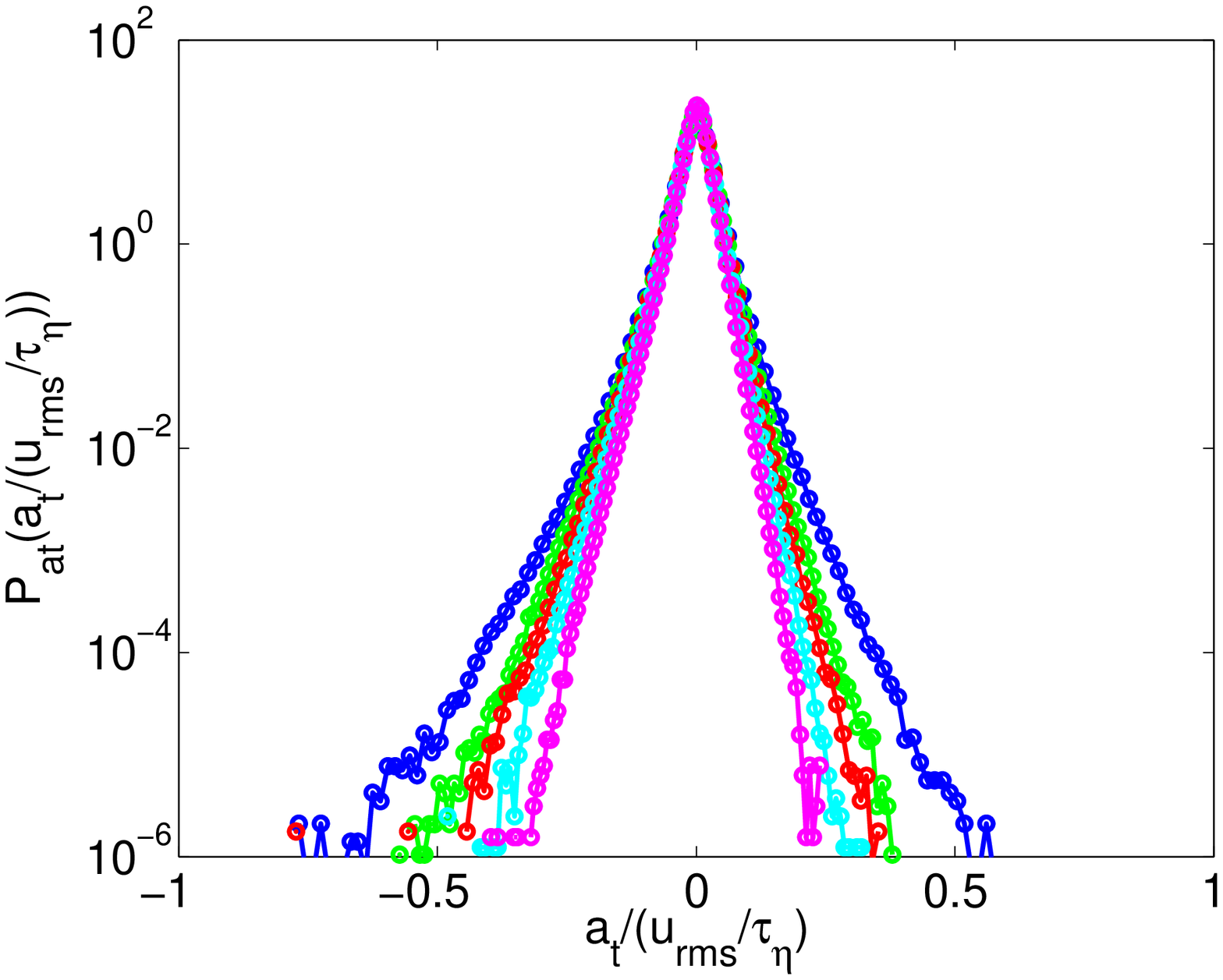}
\put(-25,100){\bf (b)}
\includegraphics[width=0.3\linewidth]{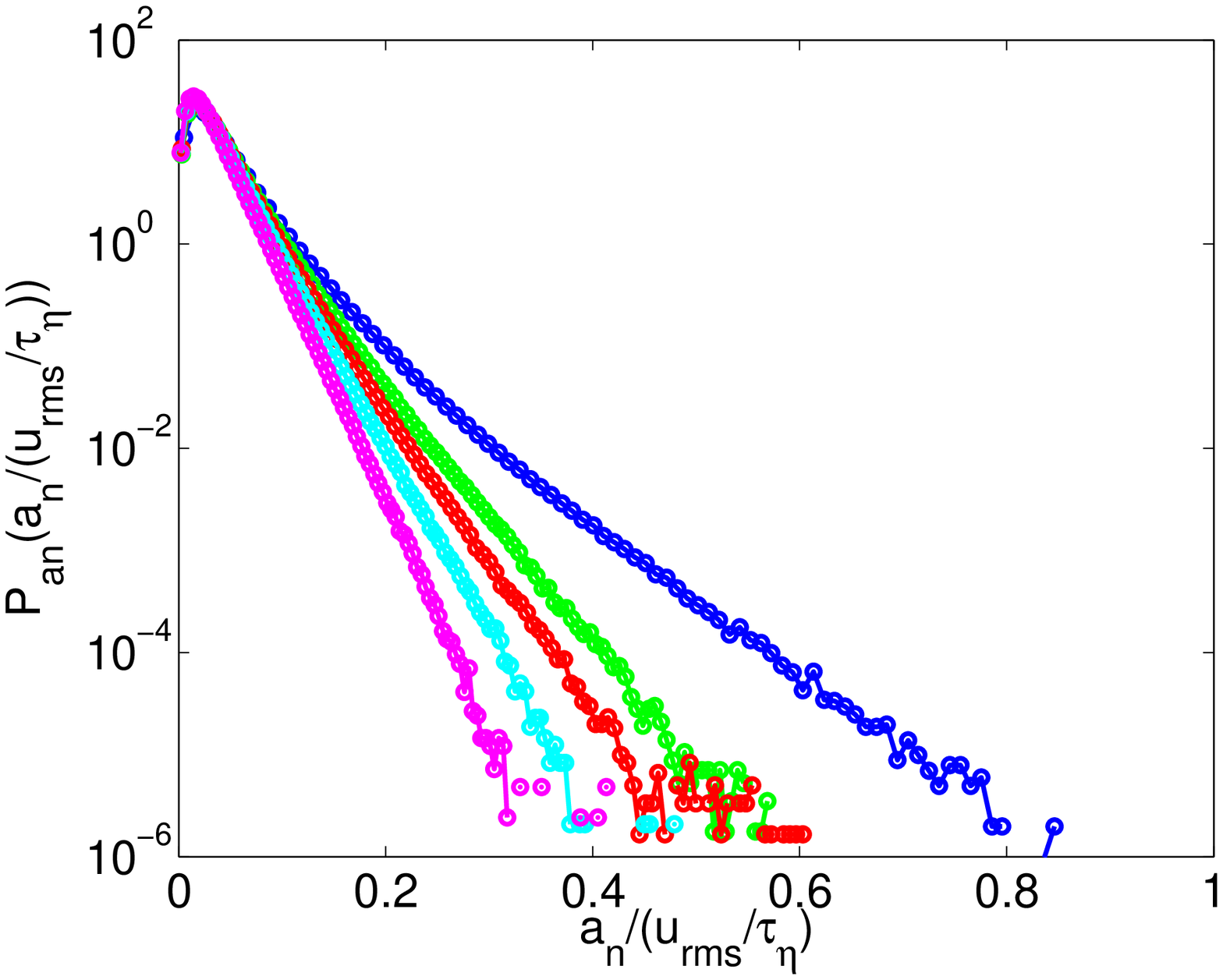}
\put(-25,100){\bf (c)}
\caption{(Color online) Probability distribution functions PDFs of (a) the modulus of 
the particle acceleration, (b) the tangential 
component of the
particle acceleration, and (c) the normal component of the particle acceleration, for $\St=0.2$
(blue curve), $\St=0.5$ (green curve), $\St=0.7$ (red curve), $\St=1.0$ (cyan curve), 
$\St=1.4$ (magenta curve), from run {\bf R2}.}
\label{fig:acc}
\end{figure*}

\begin{figure*}
\includegraphics[width=0.24\linewidth]{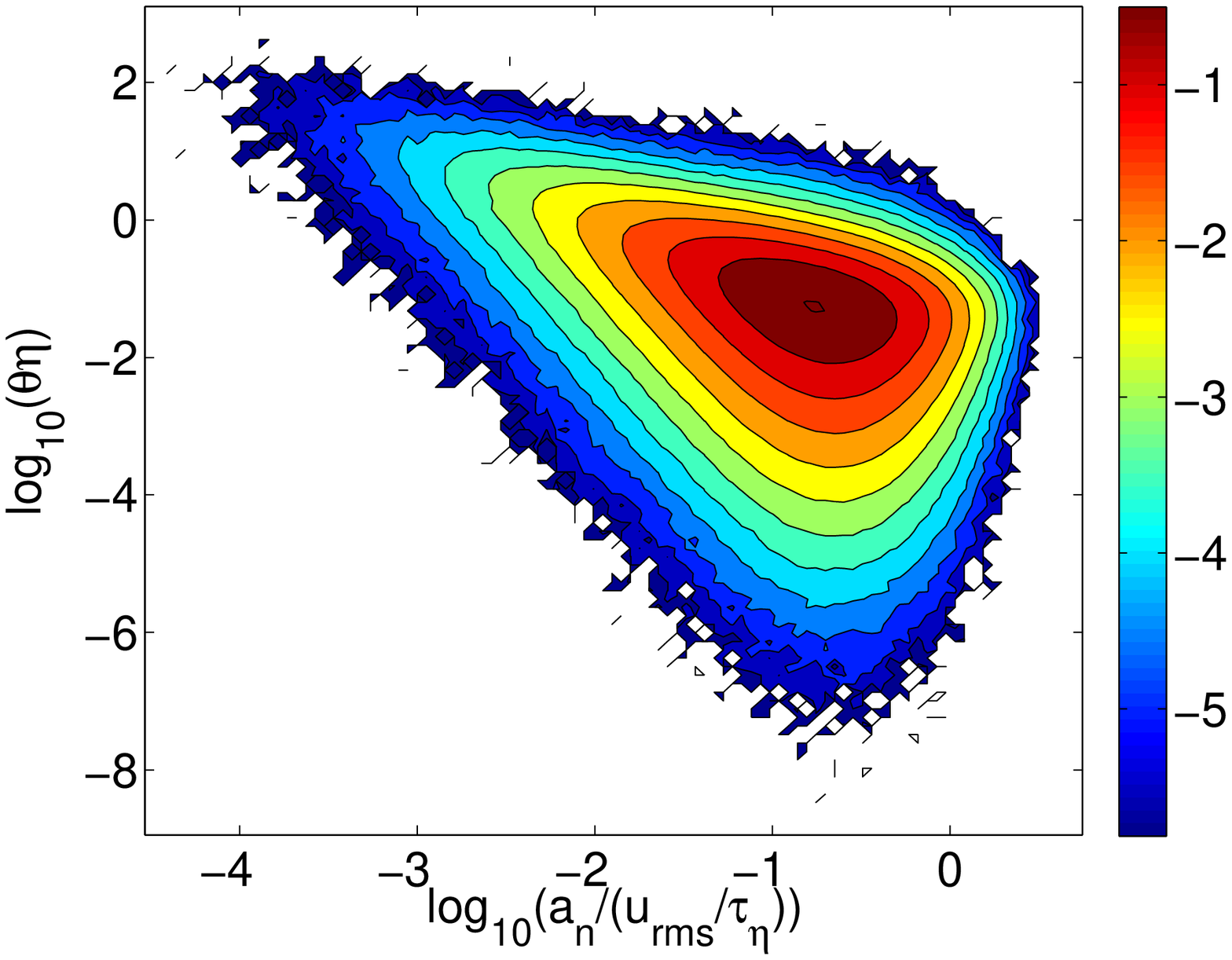}
\put(-75,95){\bf $St = 0.2$}
\put(-140,50){\bf (a)}
\includegraphics[width=0.24\linewidth]{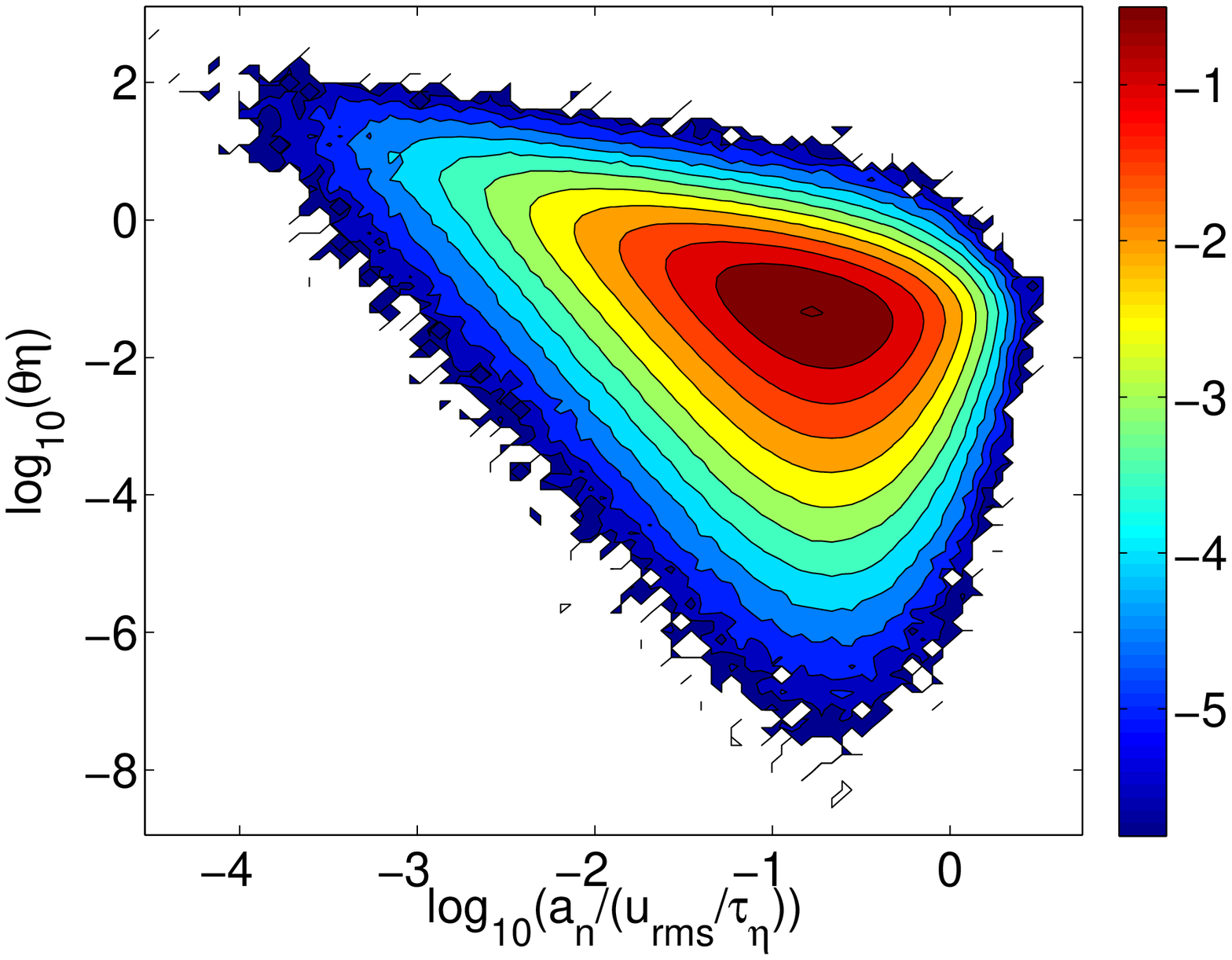}
\put(-75,95){\bf $St = 0.7$}
\includegraphics[width=0.24\linewidth]{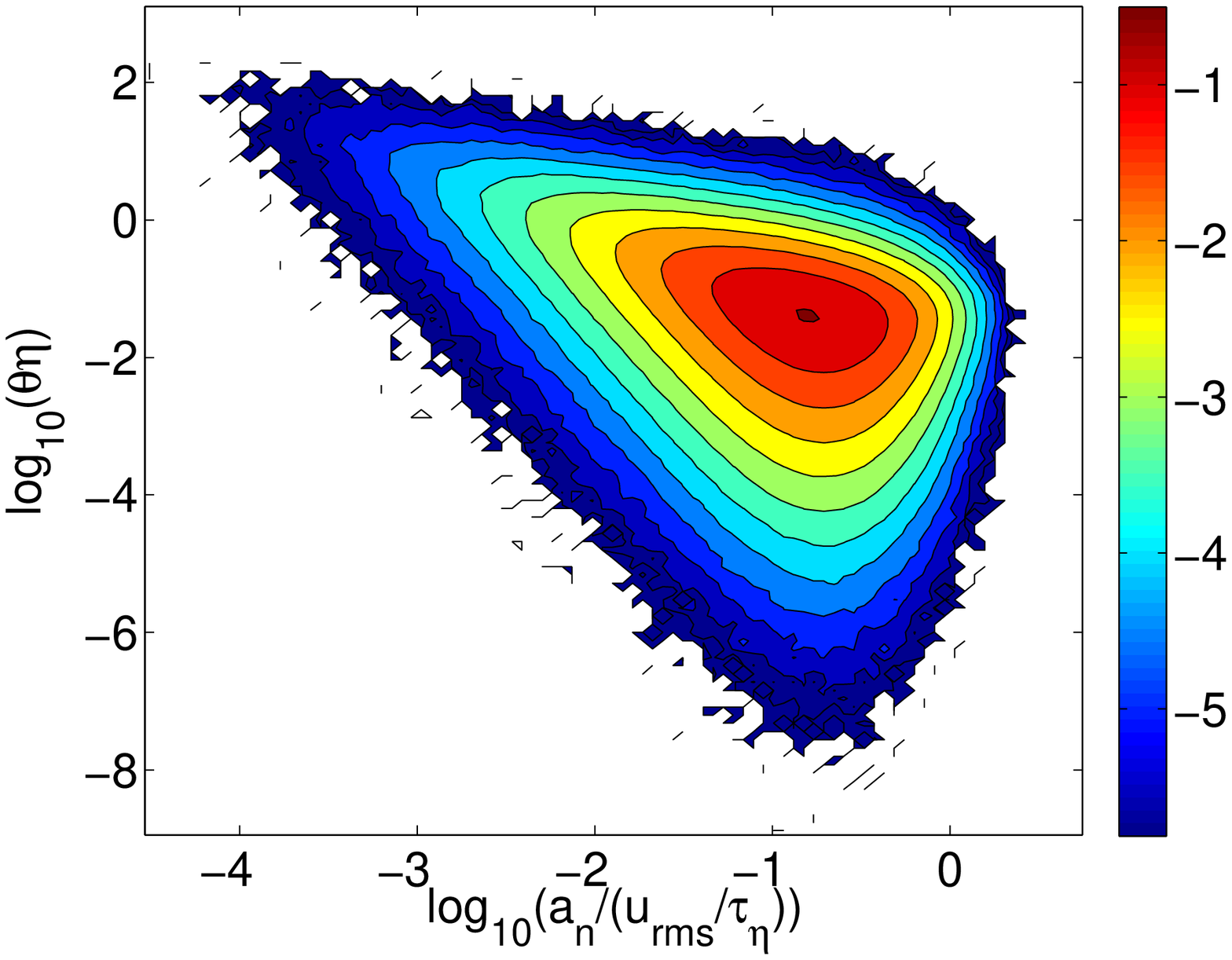}
\put(-75,95){\bf $St = 1.0$}
\includegraphics[width=0.24\linewidth]{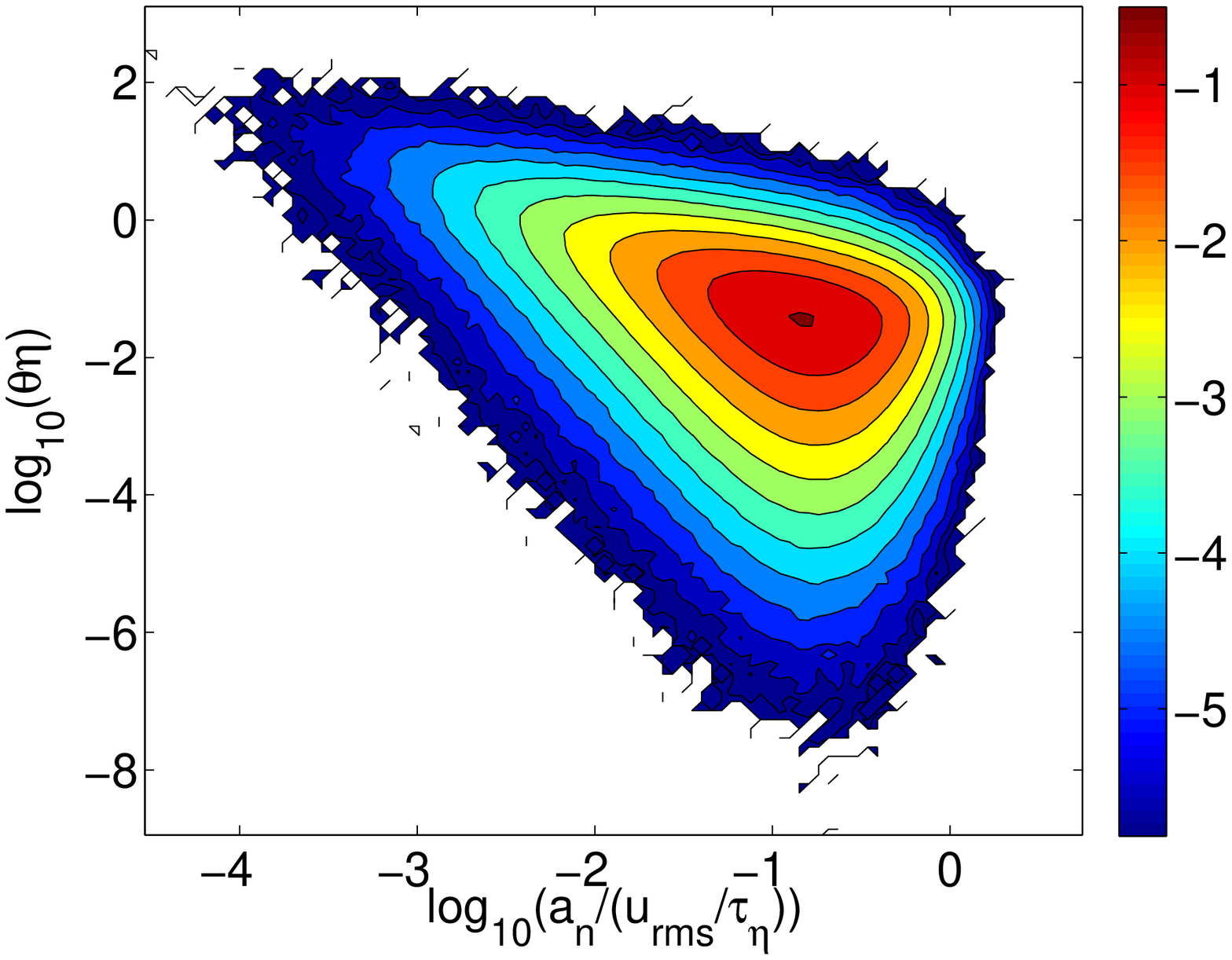}
\put(-75,95){\bf $St = 1.4$}\\

\includegraphics[width=0.24\linewidth]{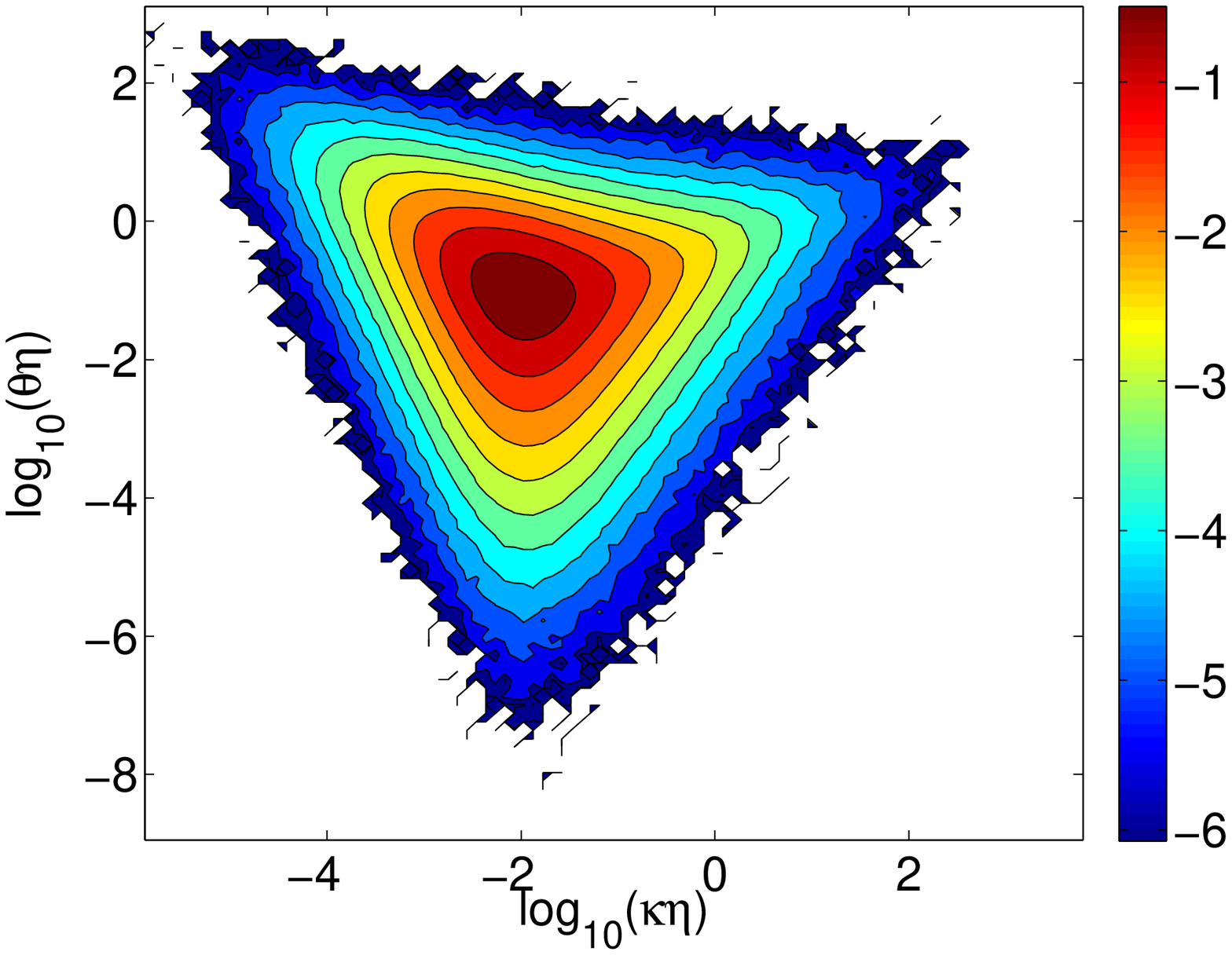}
\put(-140,50){\bf (b)}
\includegraphics[width=0.24\linewidth]{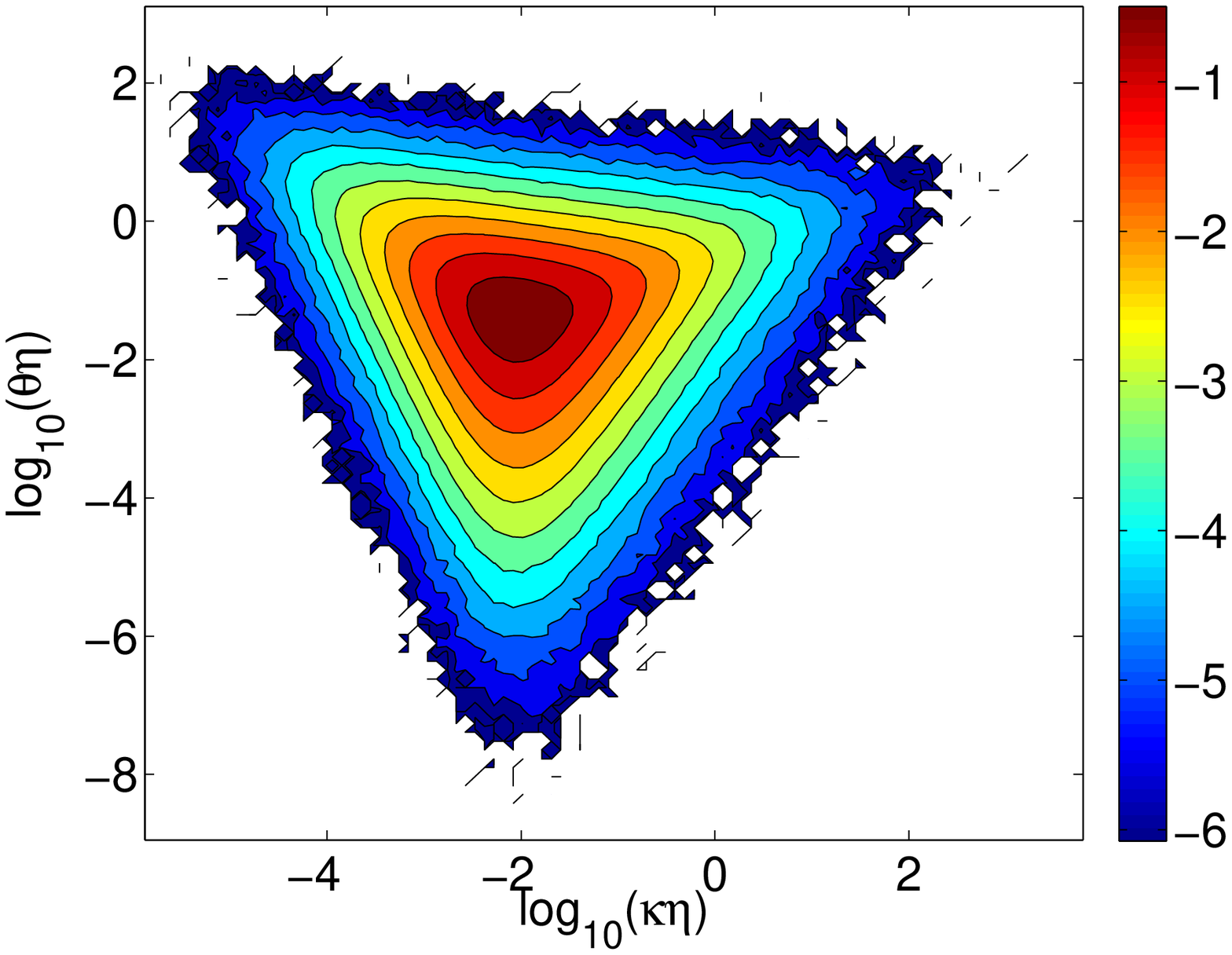}
\includegraphics[width=0.24\linewidth]{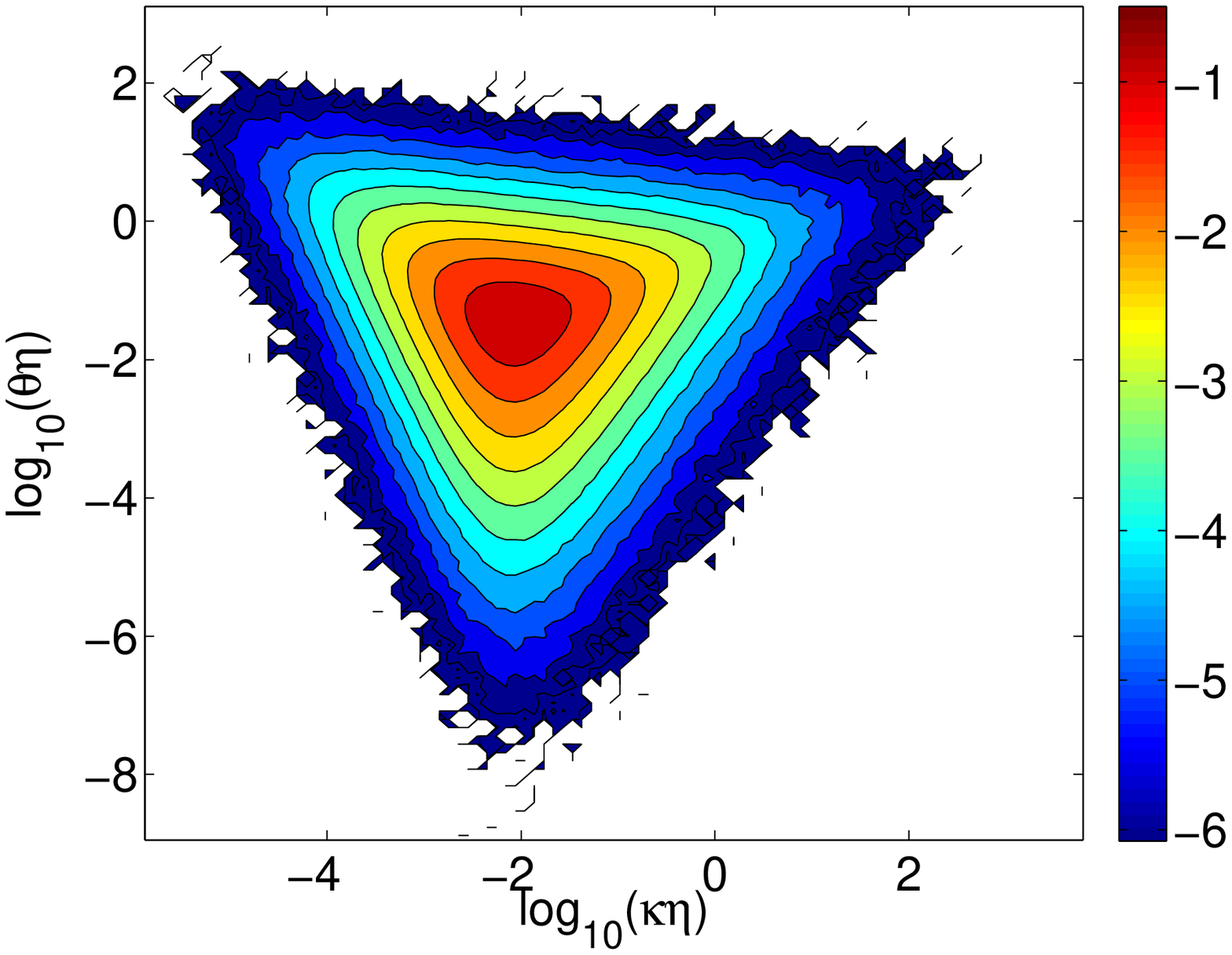}
\includegraphics[width=0.24\linewidth]{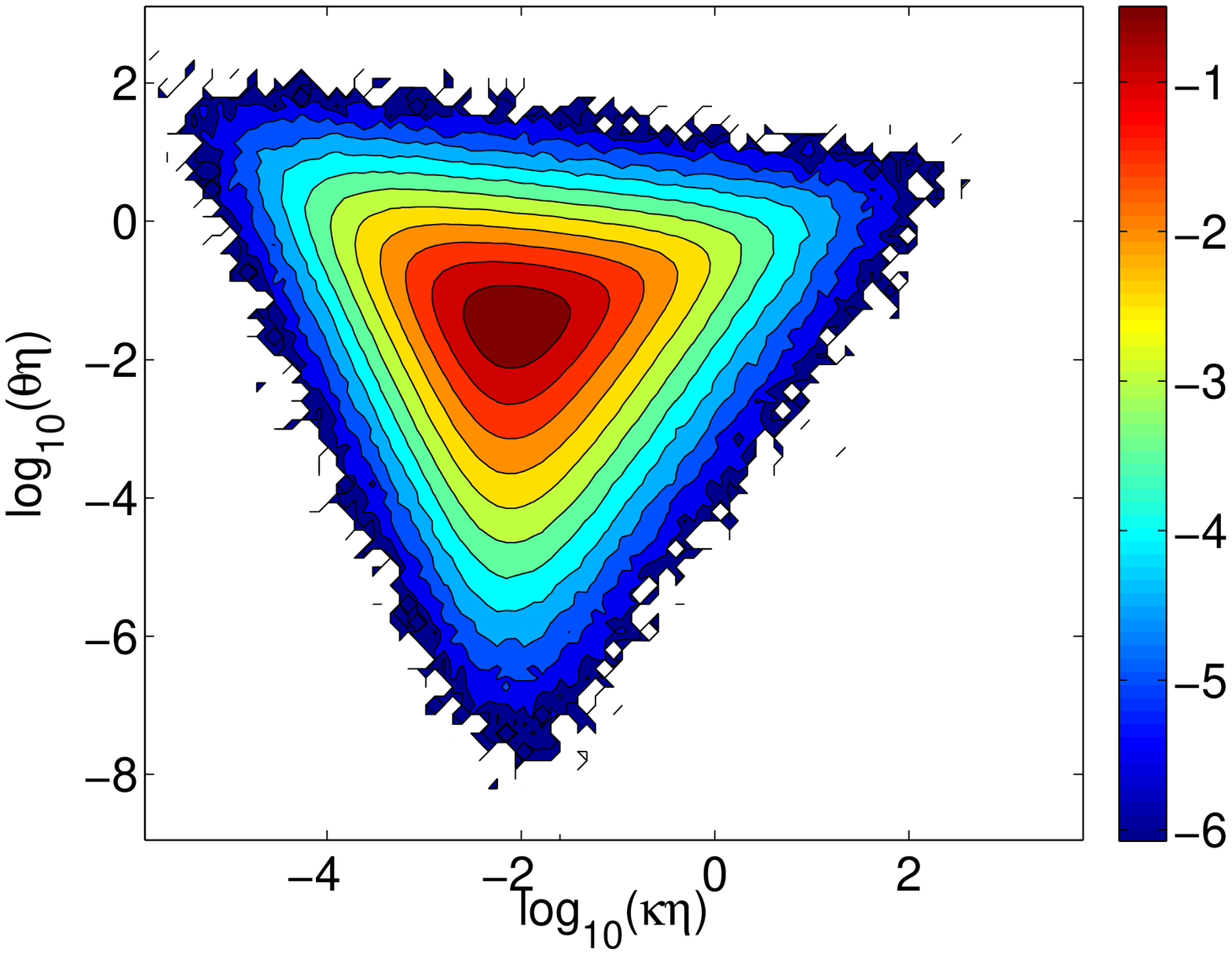} \\

\includegraphics[width=0.24\linewidth]{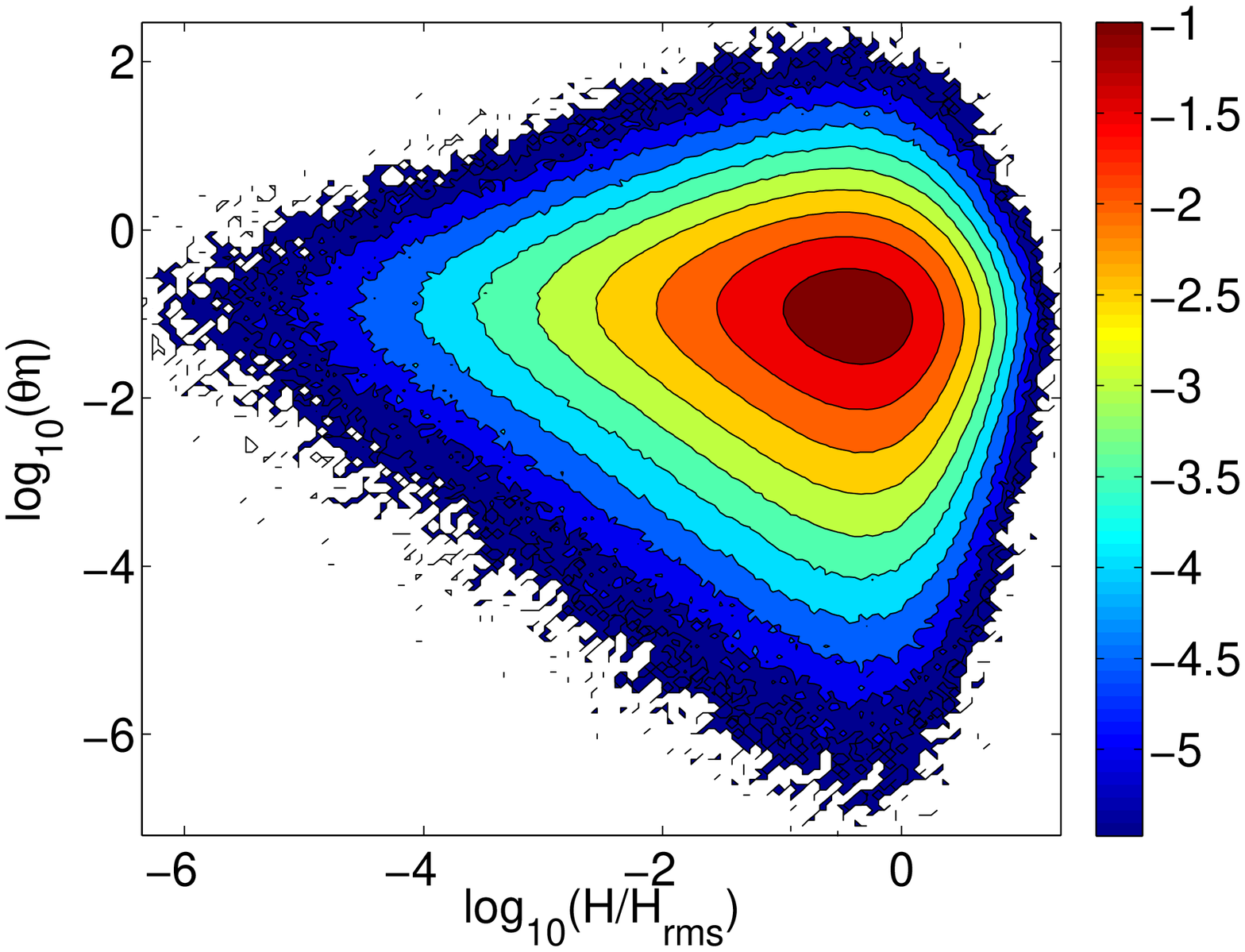}
\put(-140,50){\bf (c)}
\includegraphics[width=0.24\linewidth]{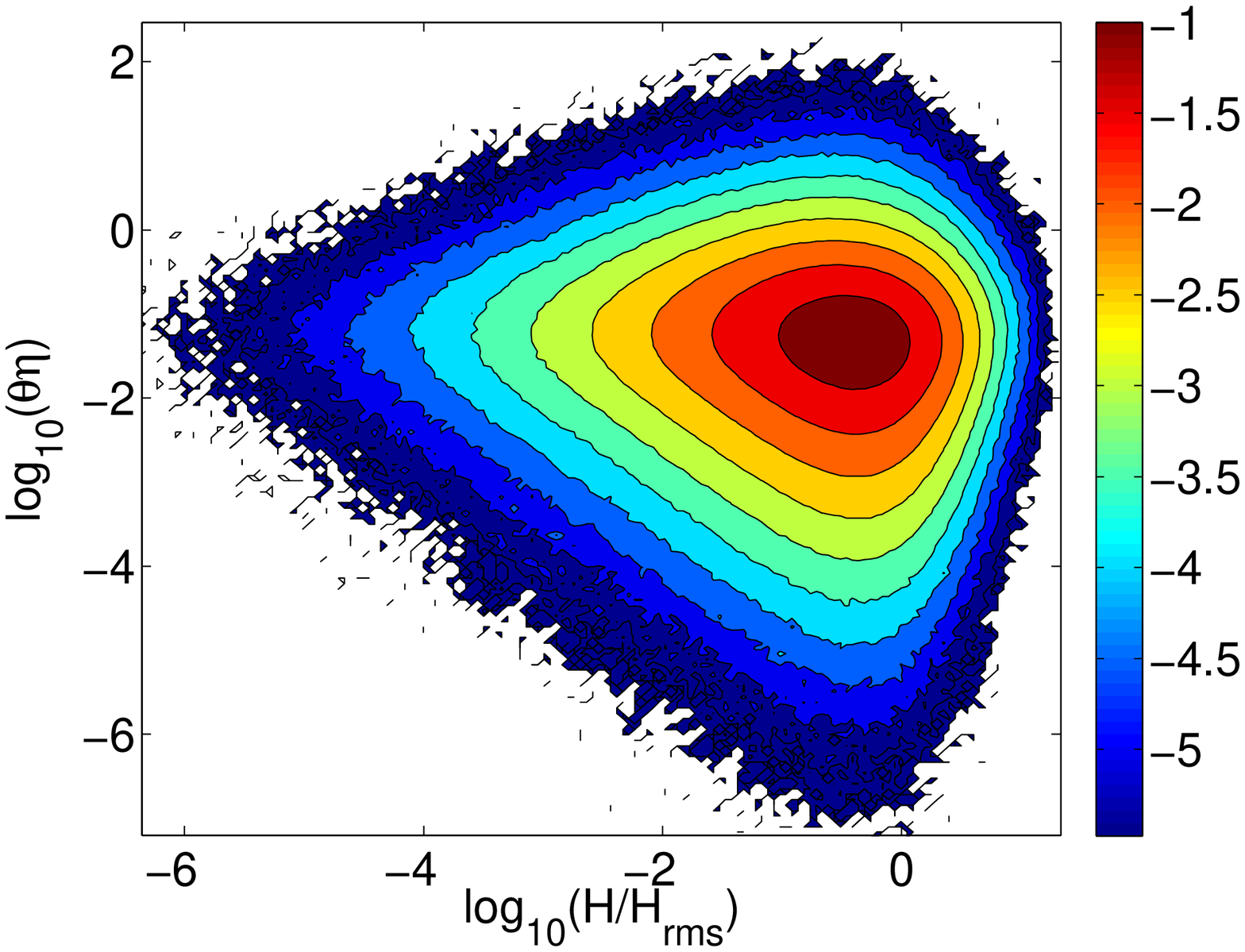}
\includegraphics[width=0.24\linewidth]{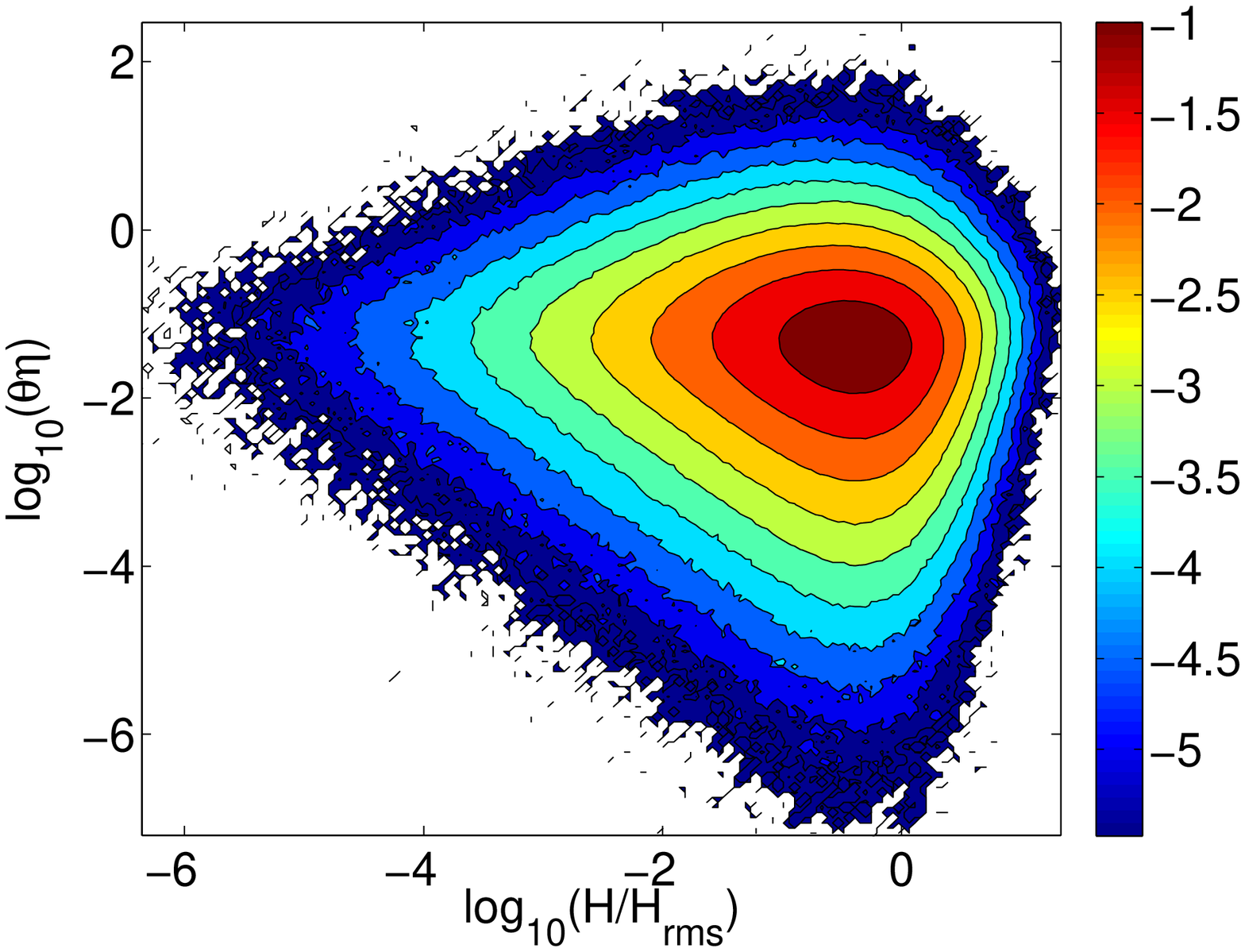}
\includegraphics[width=0.24\linewidth]{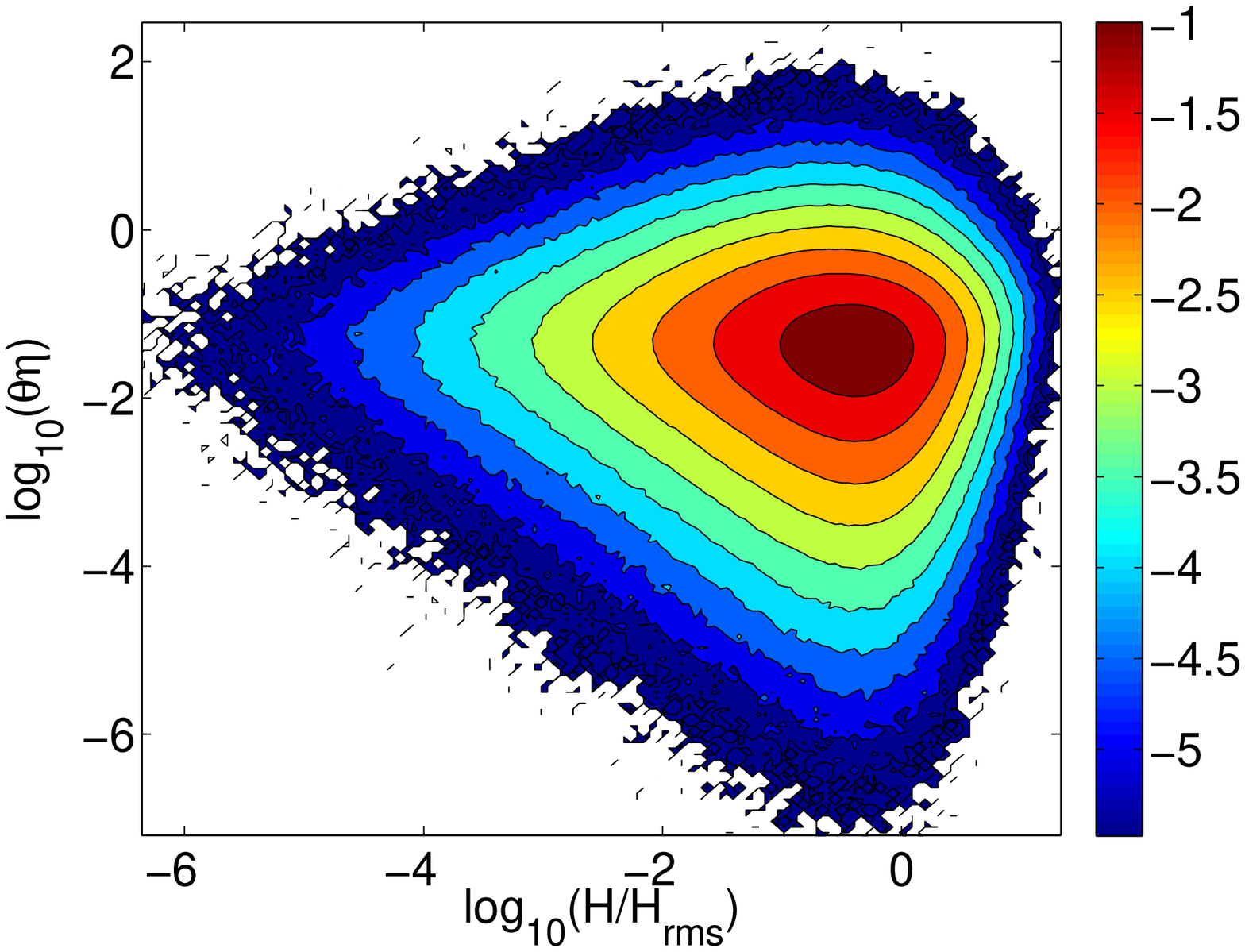} \\
\caption{(Color online) Joint PDFs of (a) the torsion and the normal component of the 
acceleration, (b) the torsion and curvature of the particle track, and (c) the torsion of 
the particle track
and the helicity of the flow at the position of the particle, for four
different values of $\St$.}
\label{fig:jpdf}
\end{figure*}

\begin{figure*}
\includegraphics[width=0.30\linewidth]{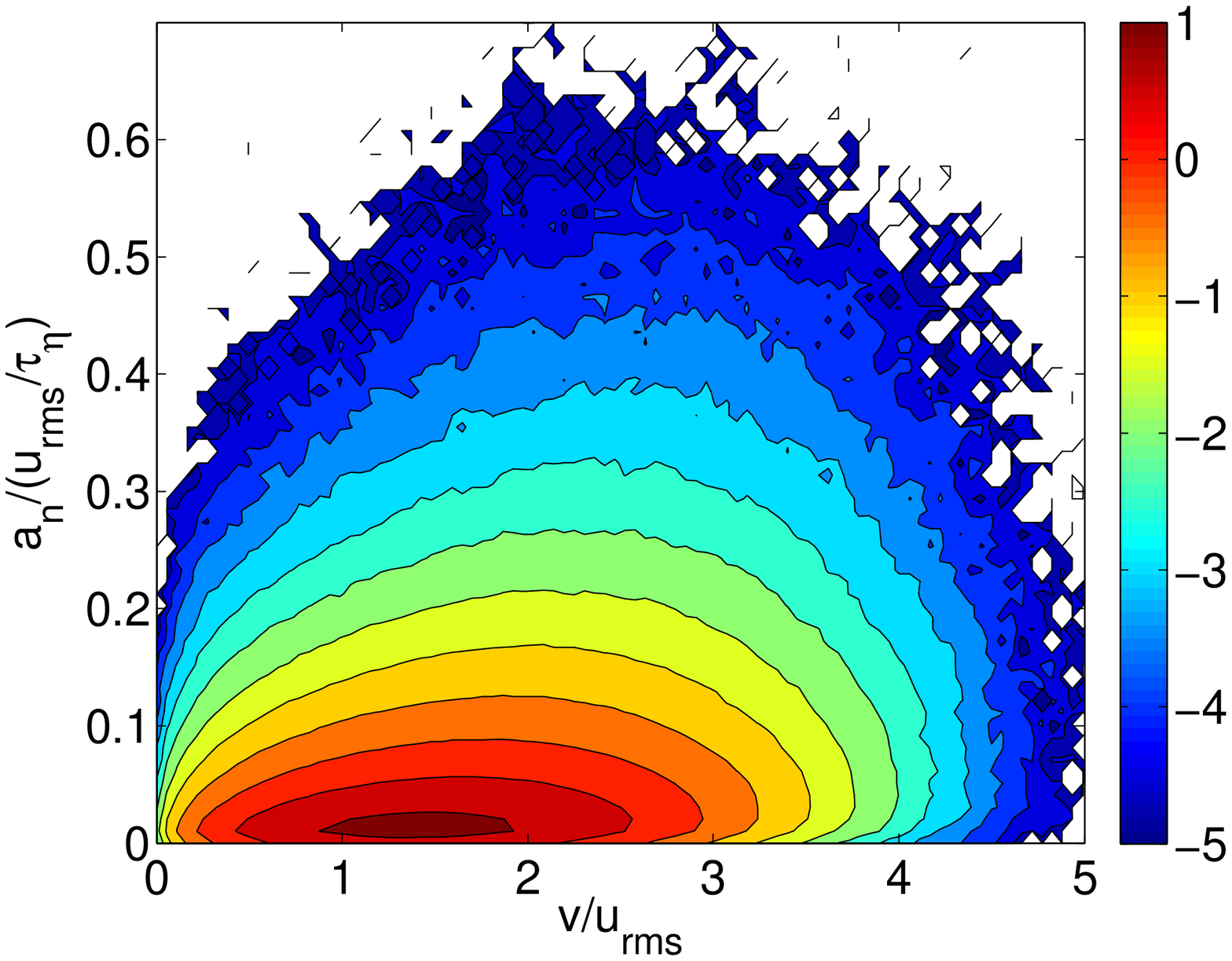}
\put(-100,130){\bf $\log_{10}{(\mathcal{P}(a_n,v))}$} 
\put(-190,100){\bf $St = 0.2$}
\includegraphics[width=0.30\linewidth]{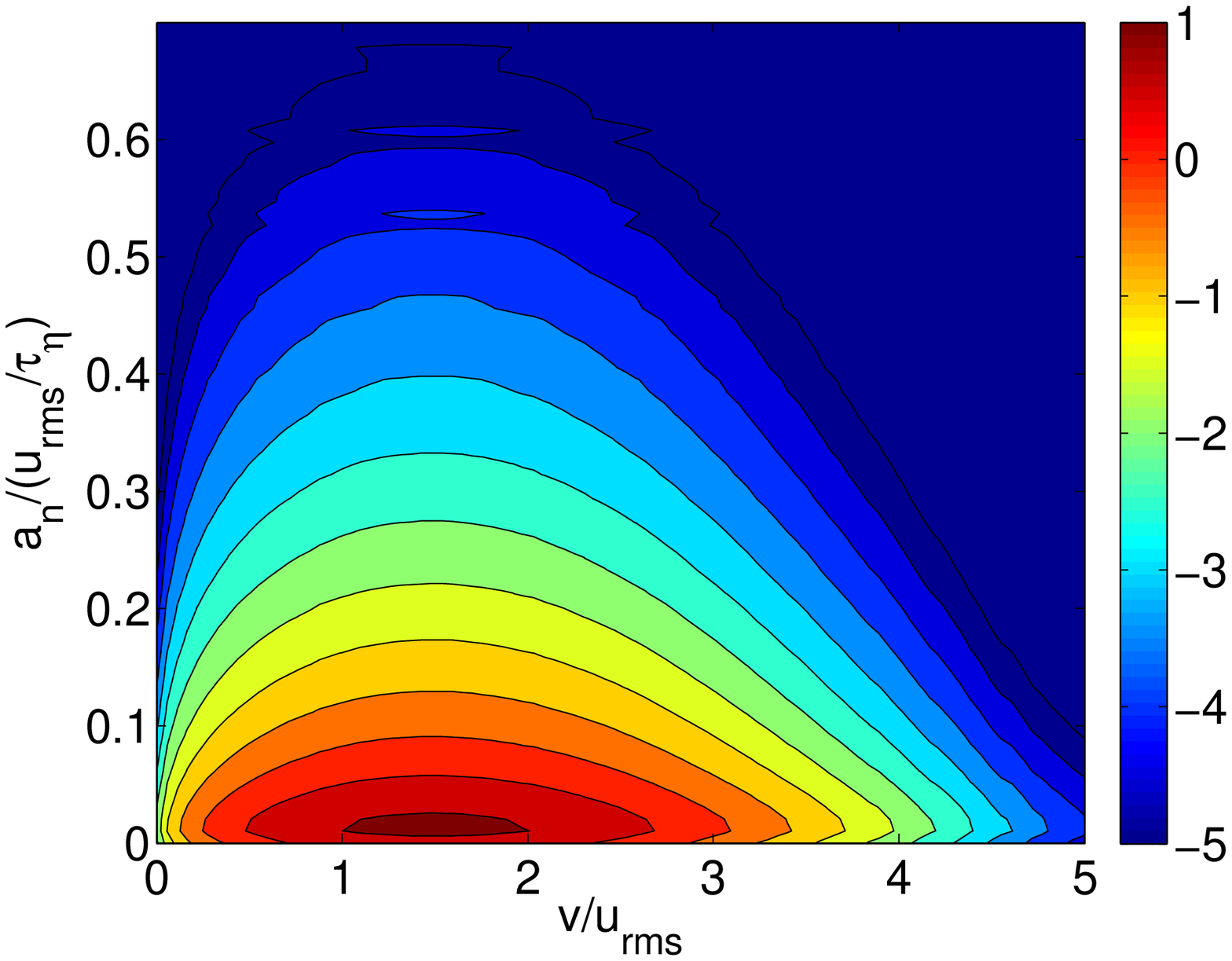}
\put(-110,130){\bf $\log_{10}{(P_{an}(a_n)P_v(v))}$} 
\includegraphics[width=0.30\linewidth]{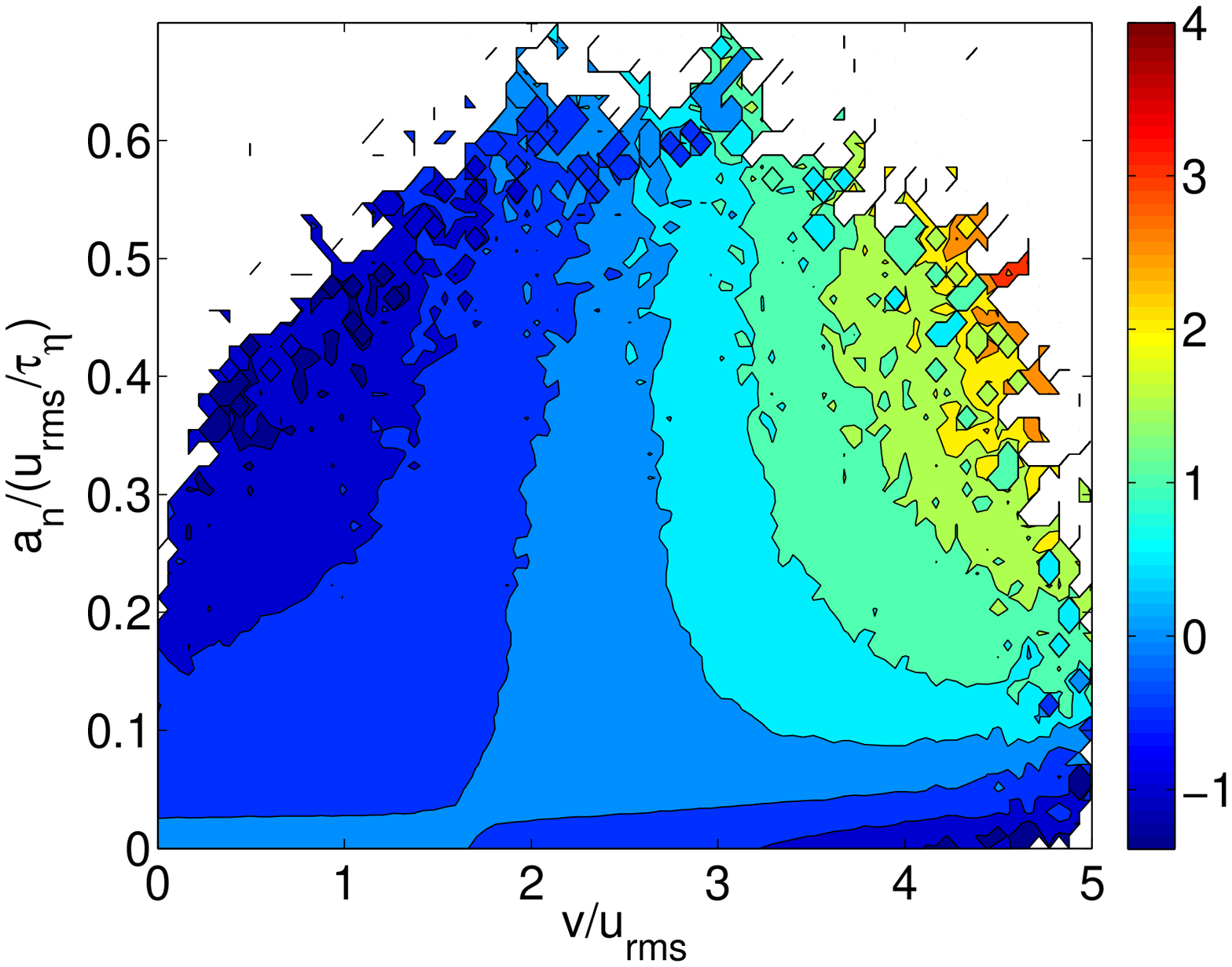} 
\put(-130,130){\bf $\log_{10}(\mathcal{P}(a_n,v)/(P_{an}(a_n) P_v(v)))$} \\
\includegraphics[width=0.30\linewidth]{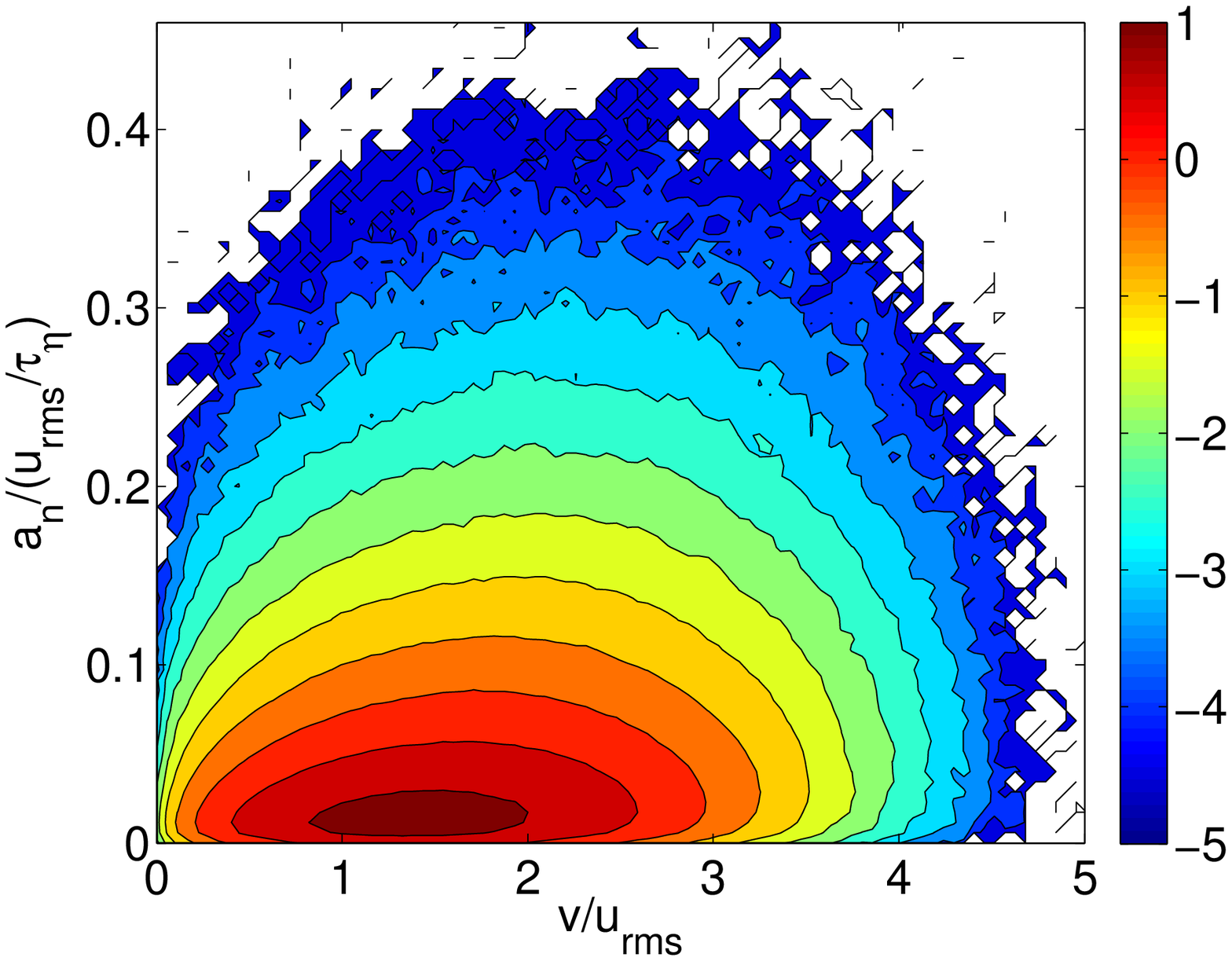}
\put(-190,100){\bf $St = 0.5$}
\includegraphics[width=0.30\linewidth]{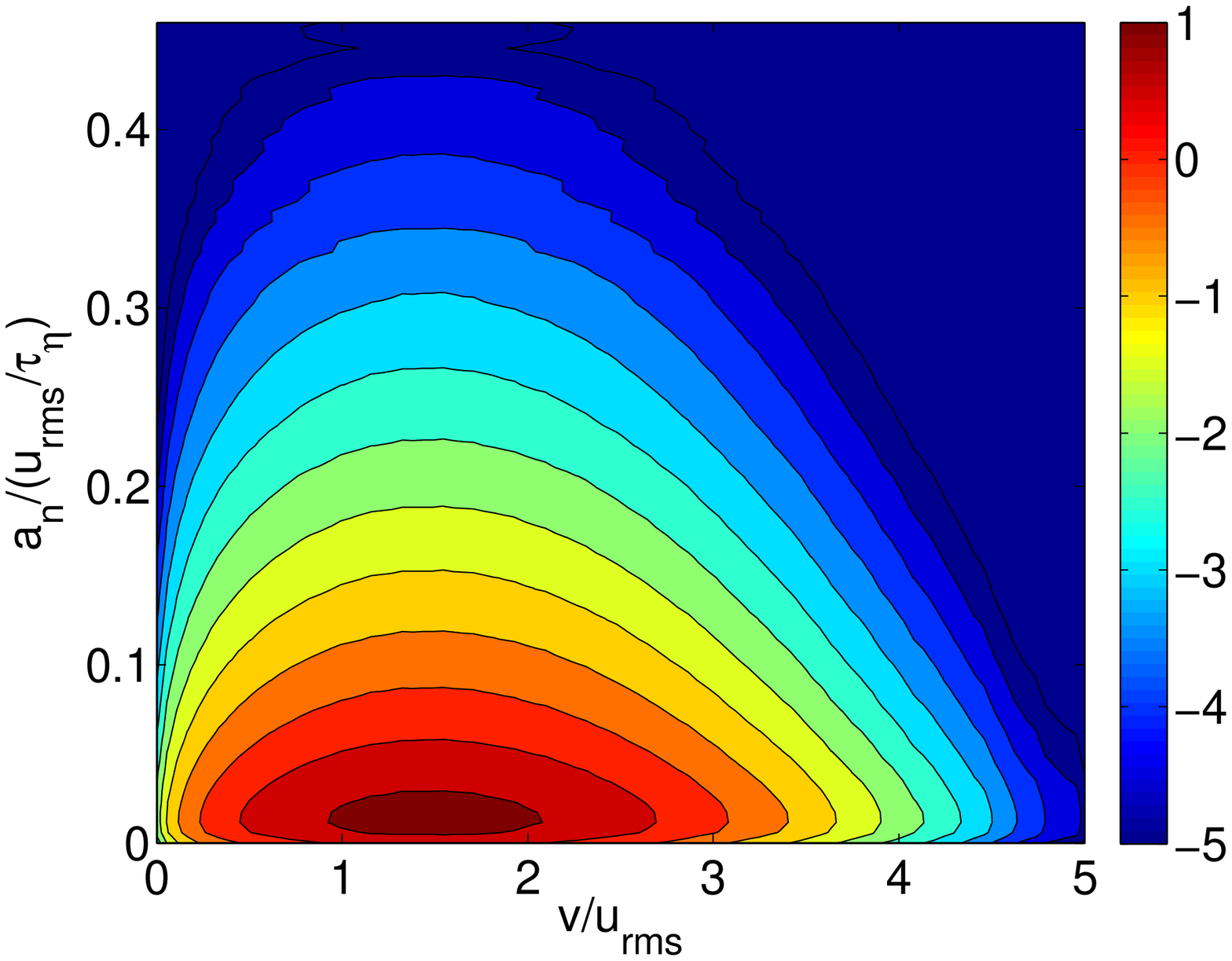}
\includegraphics[width=0.30\linewidth]{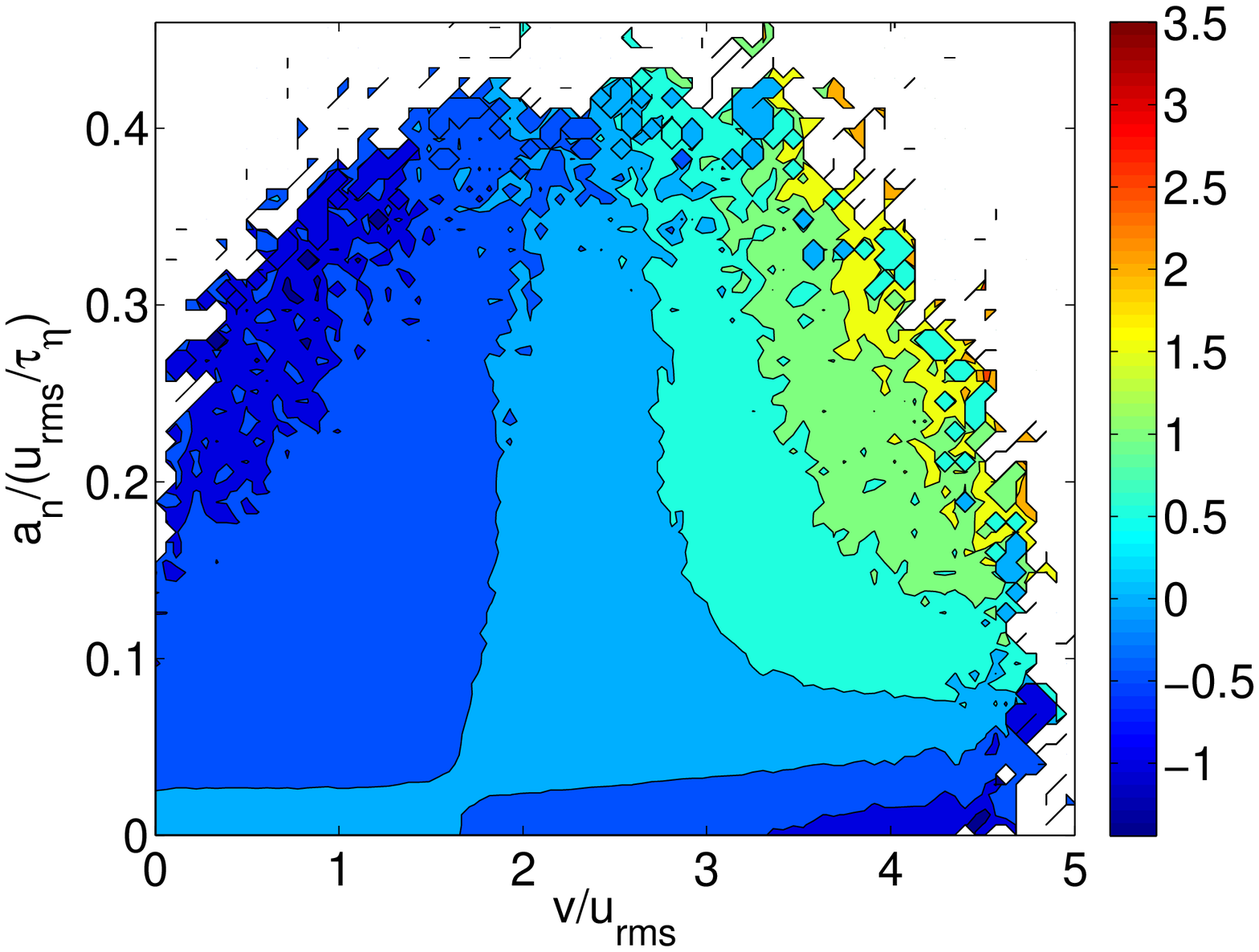} \\
\includegraphics[width=0.30\linewidth]{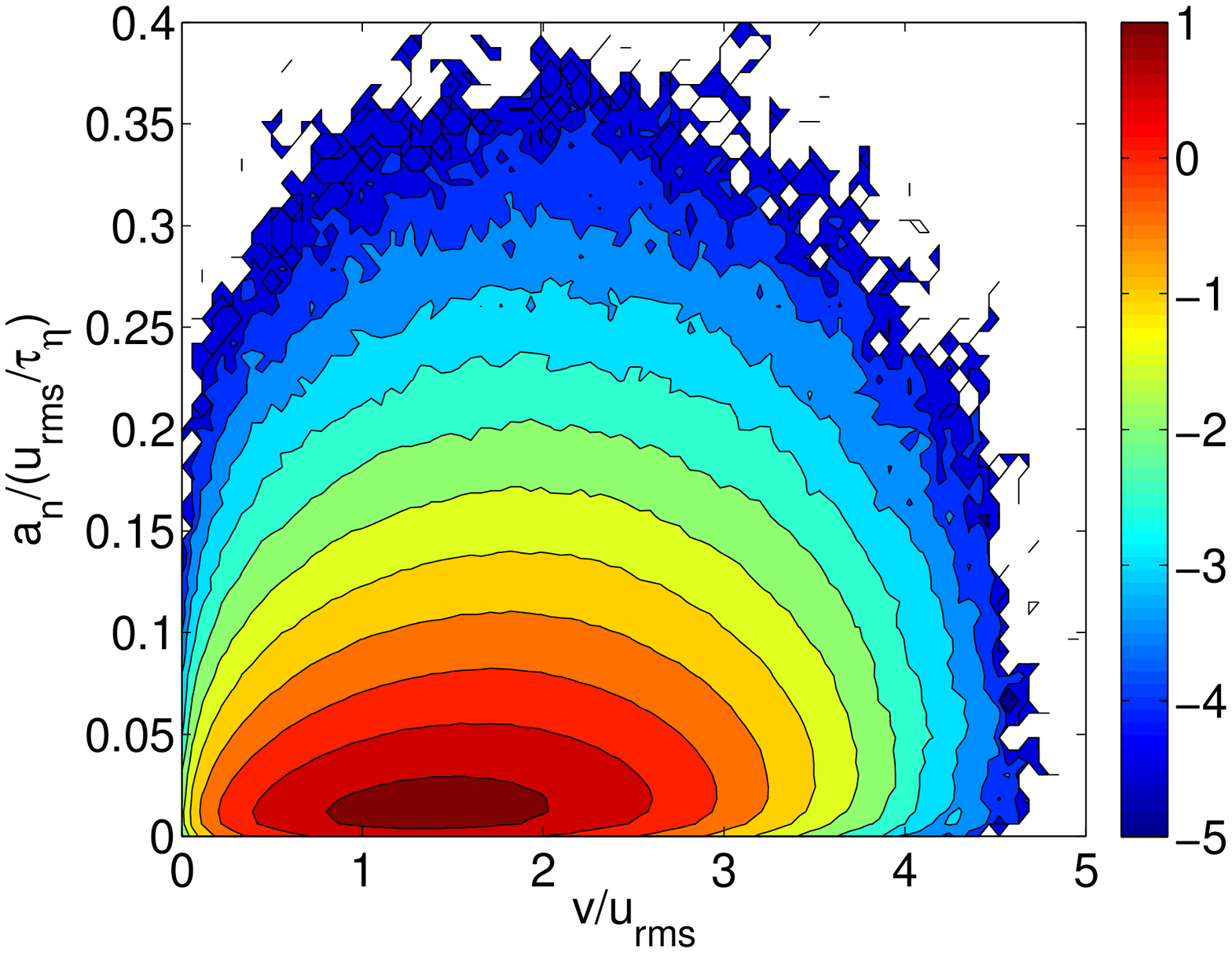}
\put(-190,100){\bf $St = 0.7$}
\includegraphics[width=0.30\linewidth]{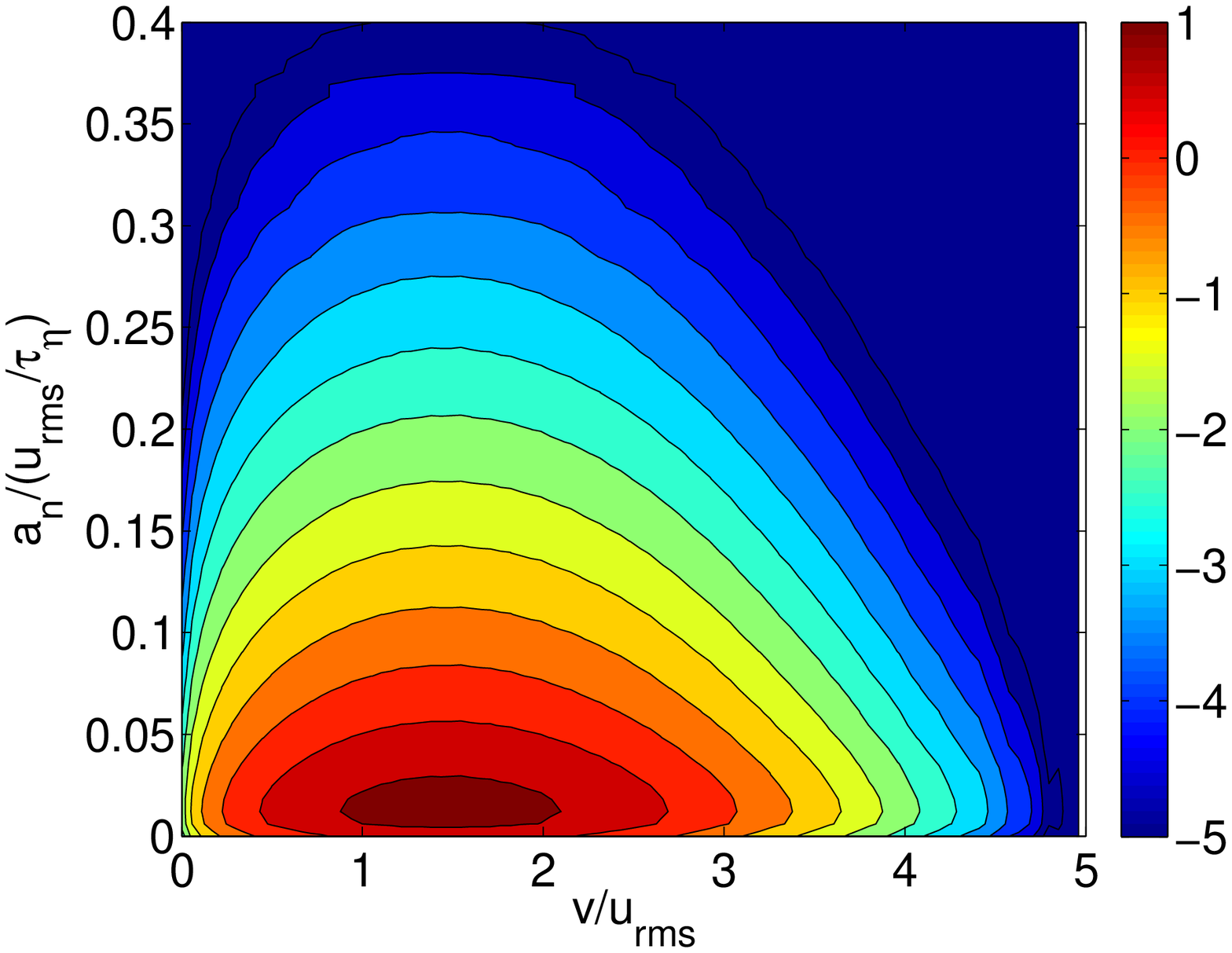}
\includegraphics[width=0.30\linewidth]{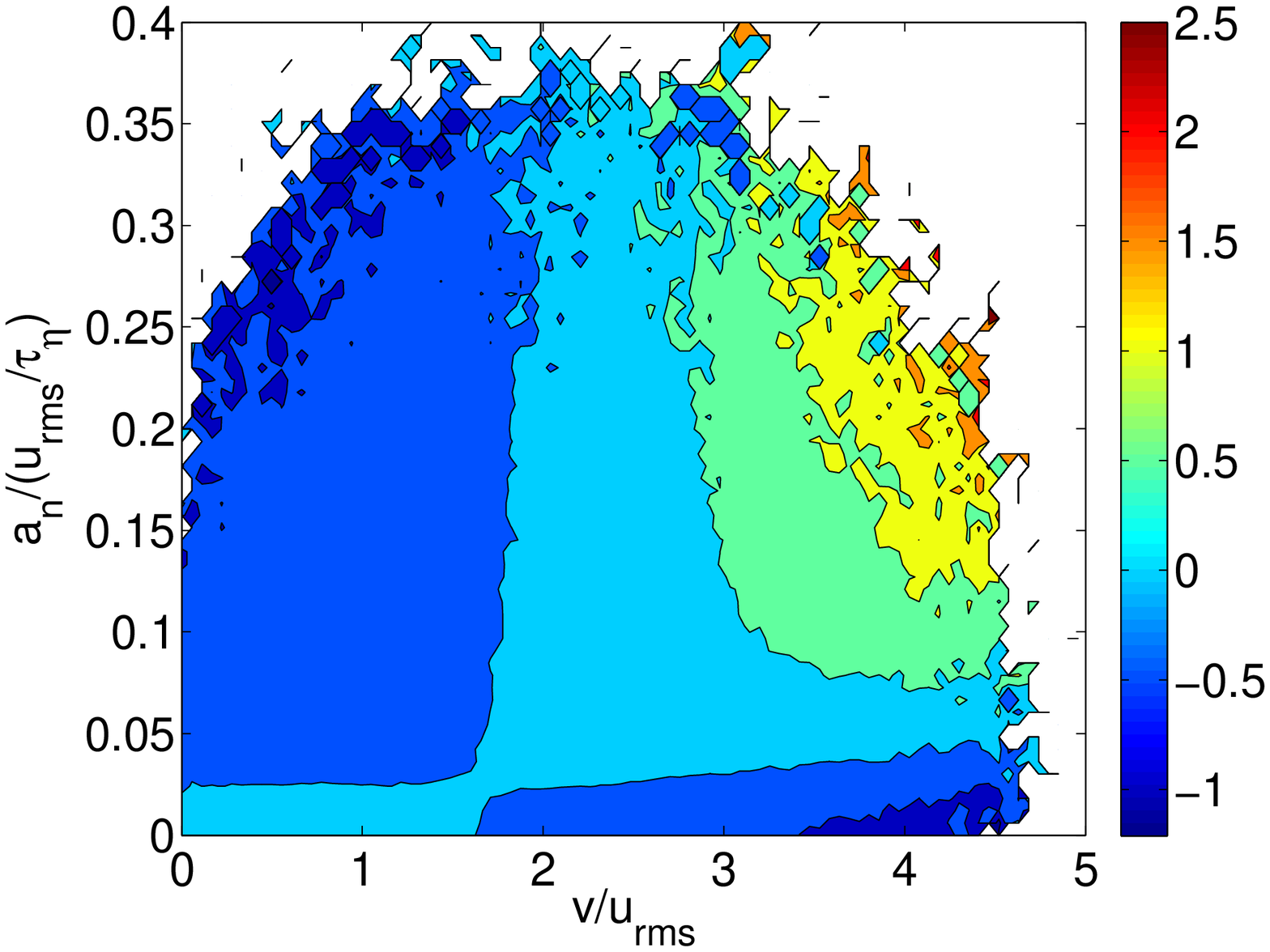} \\
\includegraphics[width=0.30\linewidth]{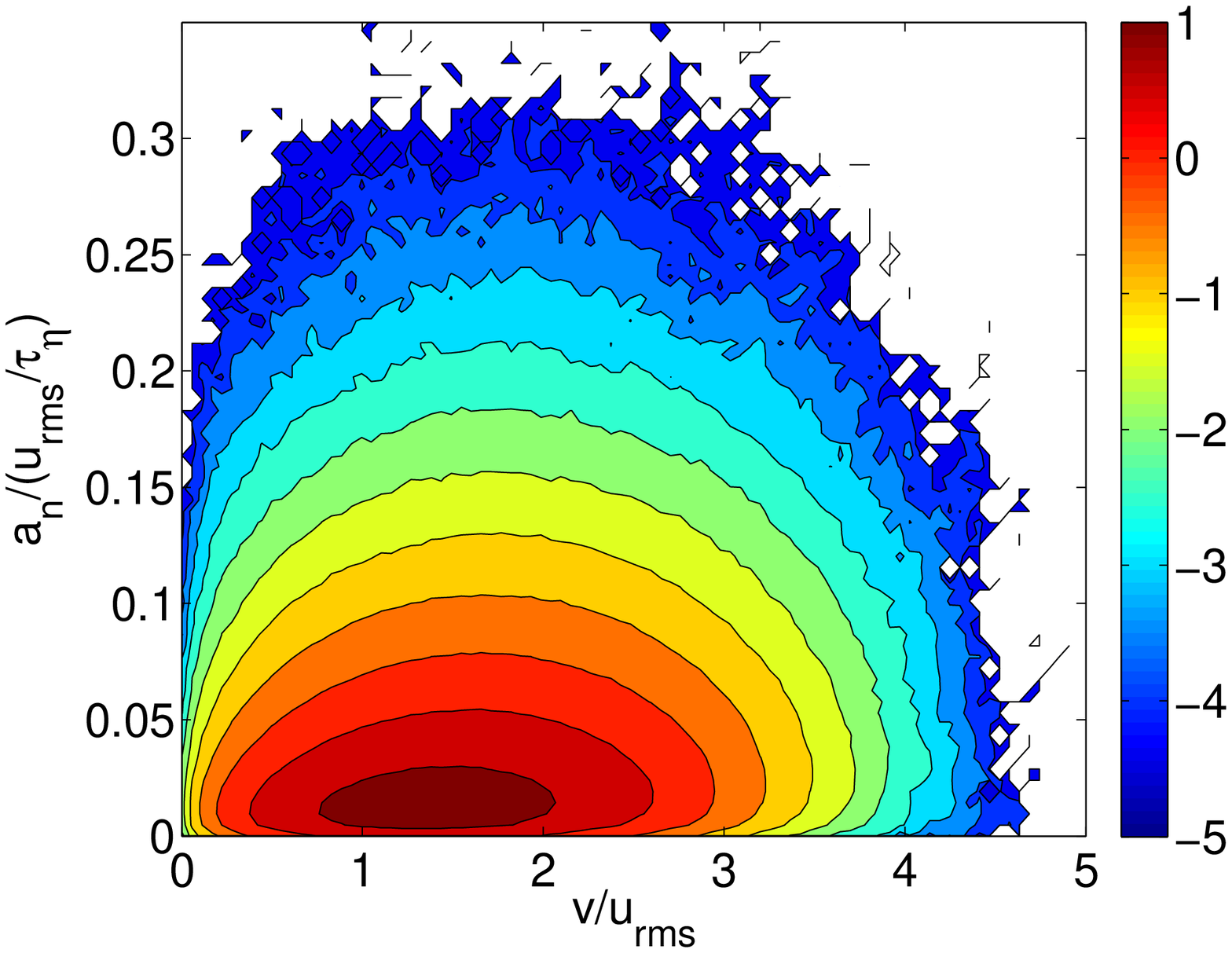}
\put(-190,100){\bf $St = 1.0$}
\includegraphics[width=0.30\linewidth]{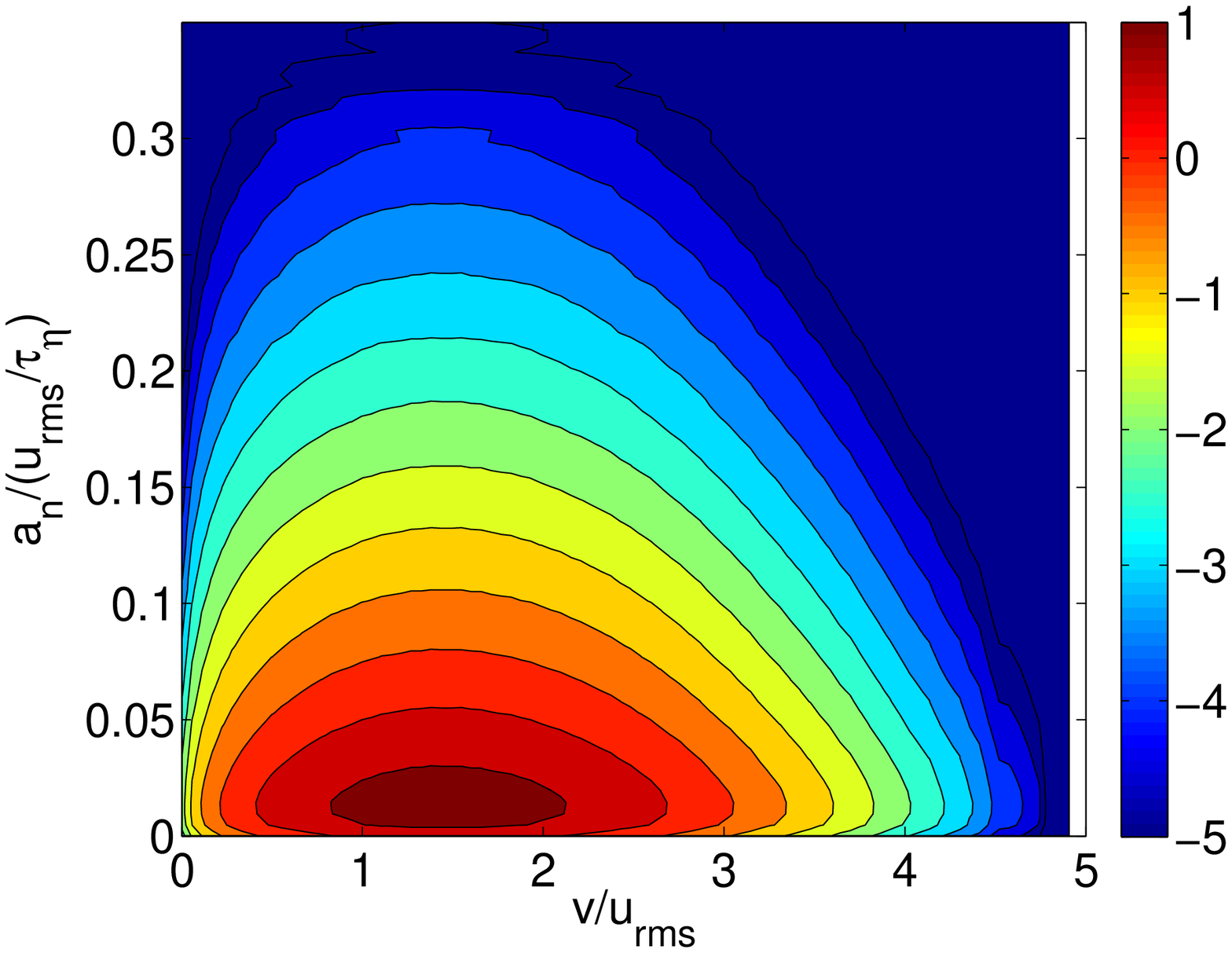}
\includegraphics[width=0.30\linewidth]{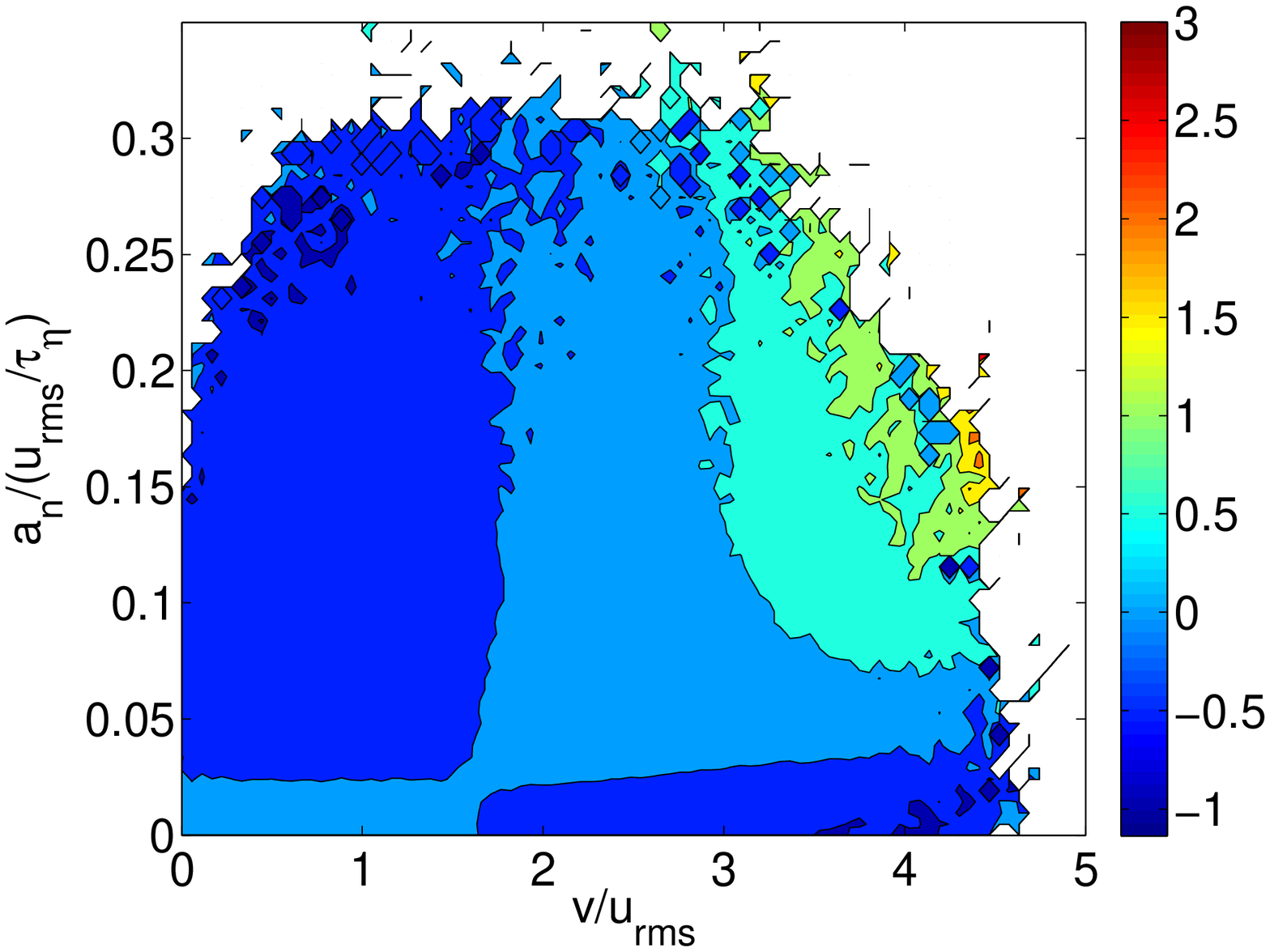} \\ 
\includegraphics[width=0.30\linewidth]{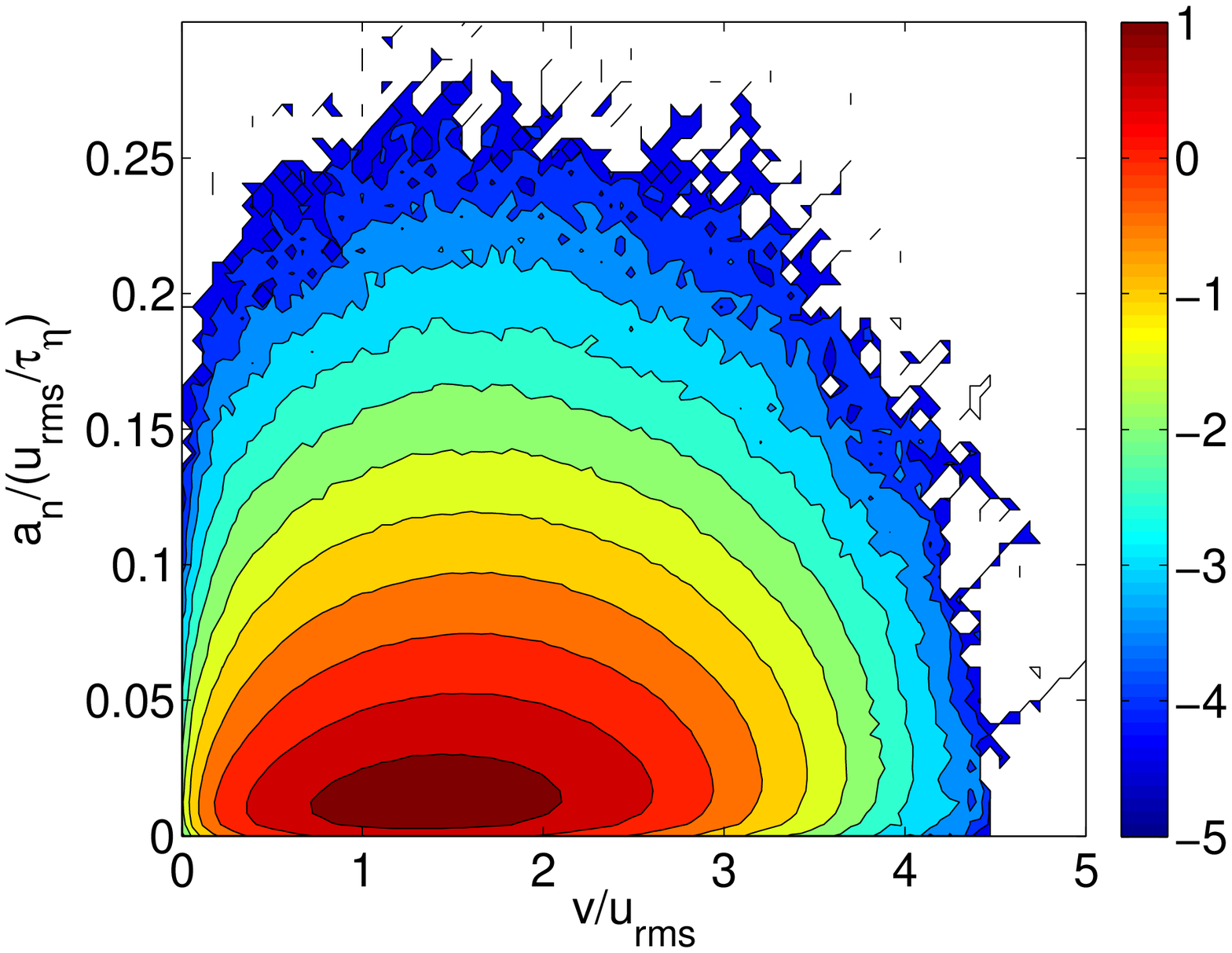}
\put(-190,100){\bf $St = 1.4$}
\includegraphics[width=0.30\linewidth]{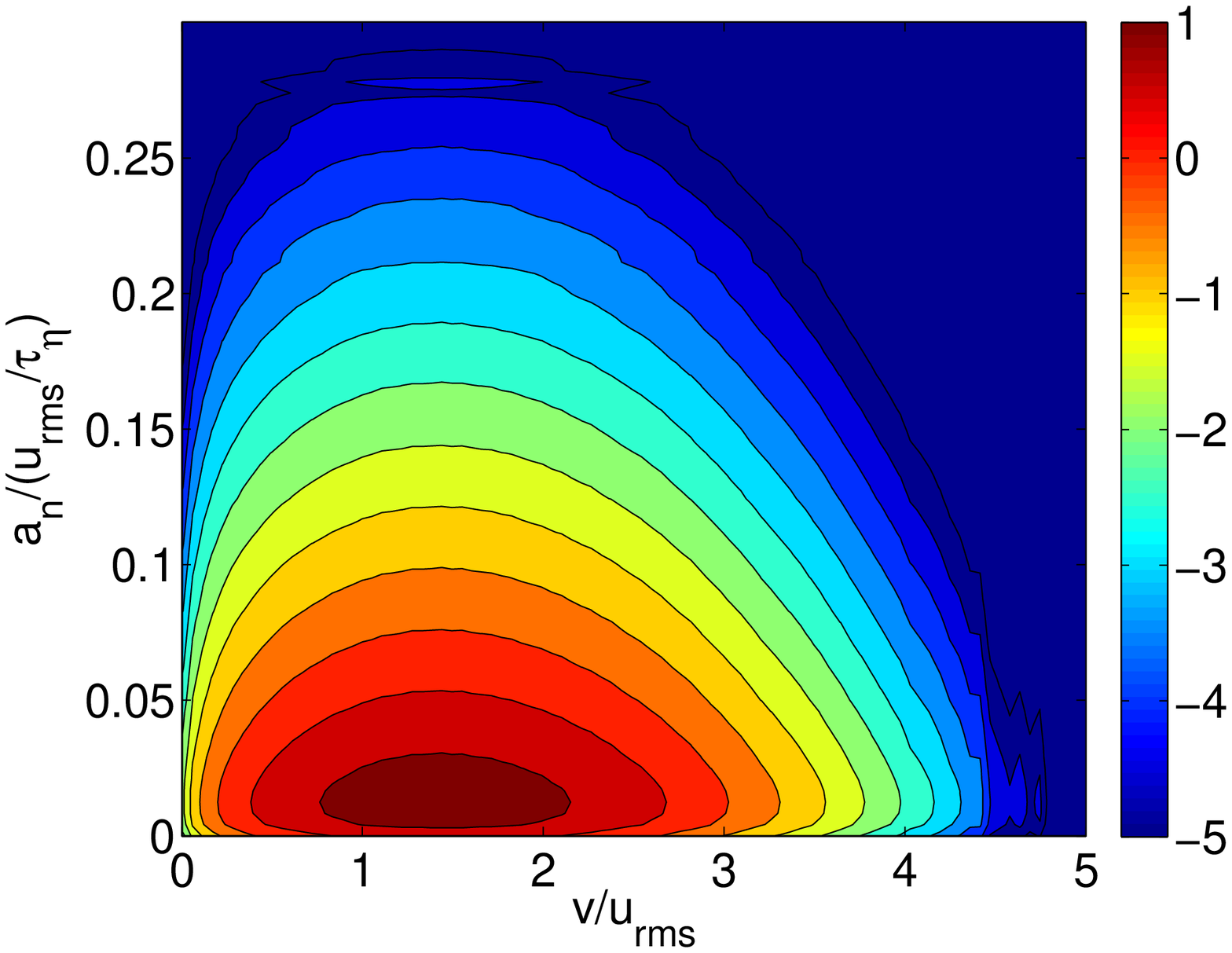}
\includegraphics[width=0.30\linewidth]{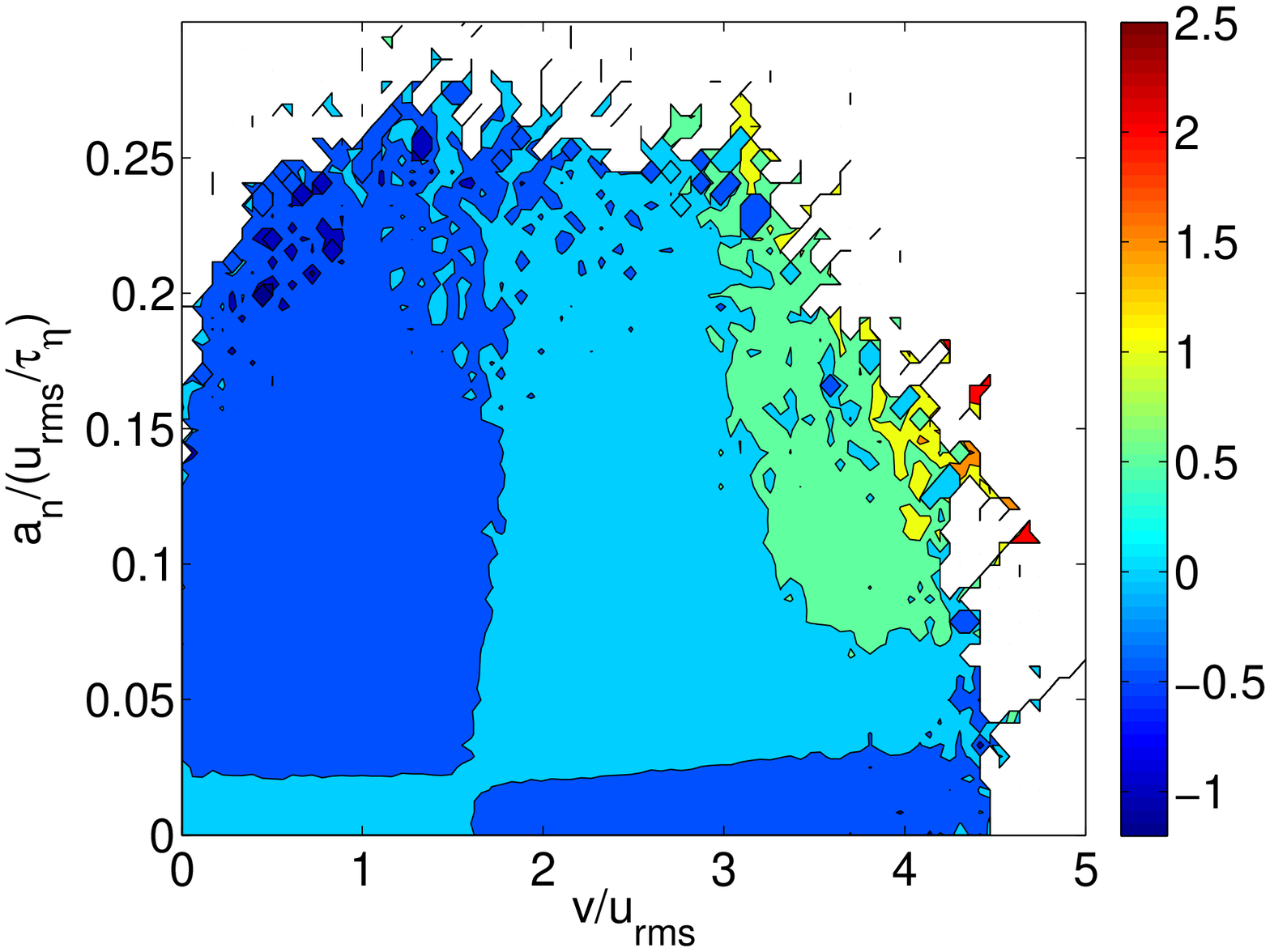} 
\caption{(Color online) Contours of the joint PDFs $\mathcal{P}(a_n,v)$ (left column), 
 the product of the
independent PDFs $P_{an}(a_n) P_v(v)$ (middle column), and  
$\log_{10}(\mathcal{P}(a_n,v)/(P_{an}(a_n) P_v(v)))$ (right column).}
\label{fig:jpdfanv}
\end{figure*}

\begin{figure*}
\includegraphics[width=0.30\linewidth]{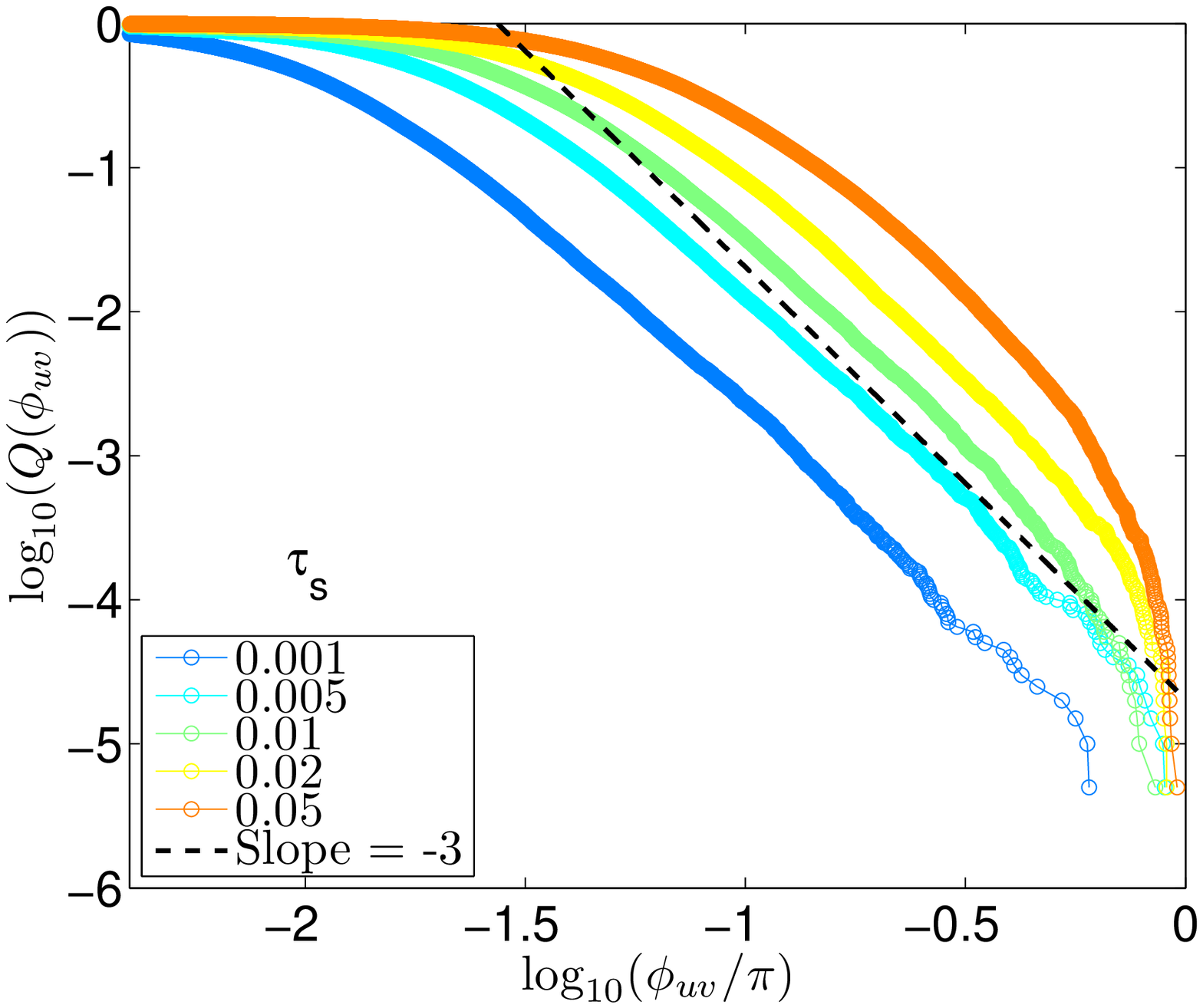}          
\put(-25,100){\bf (a)}
\includegraphics[width=0.30\linewidth]{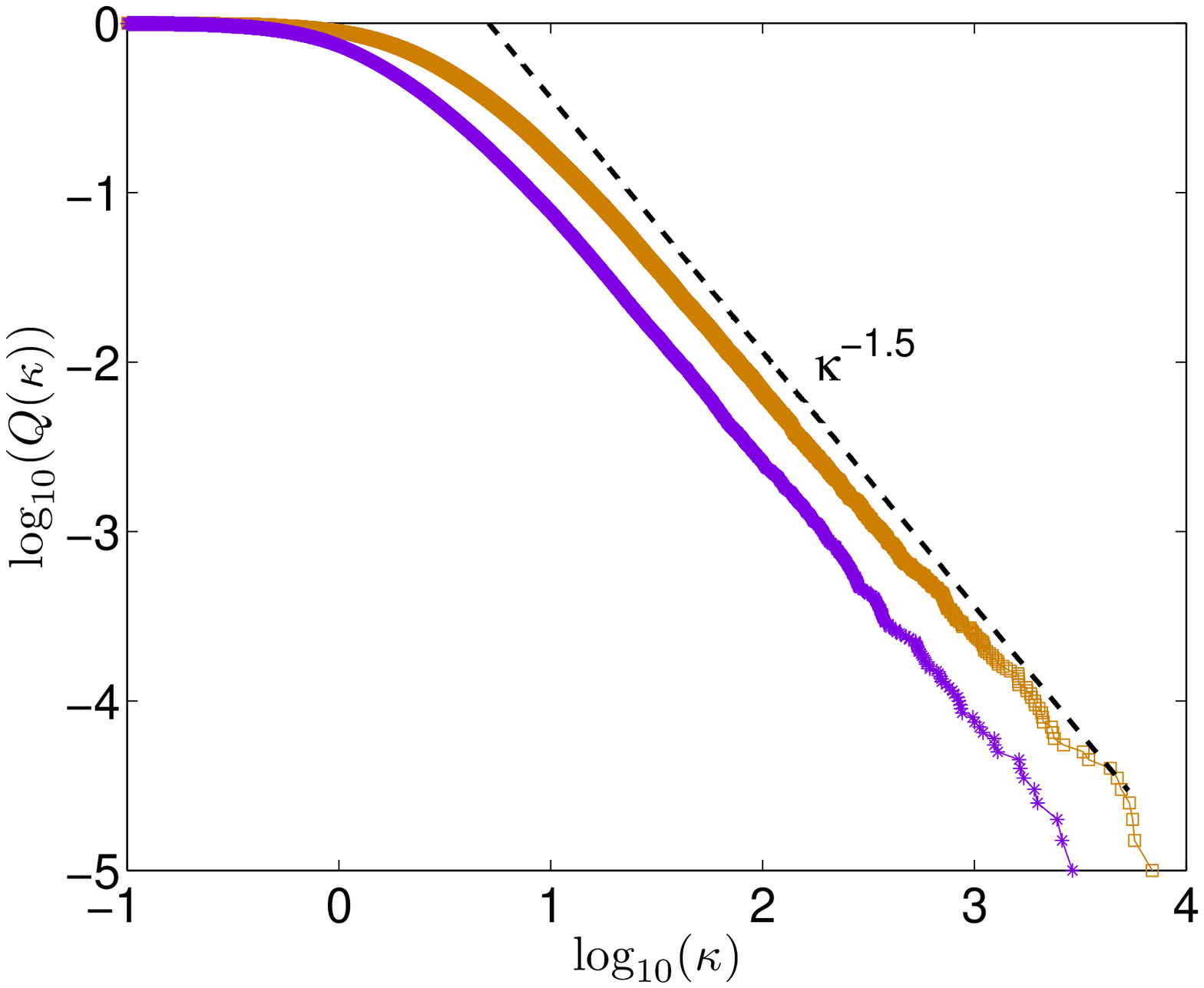}          
\put(-25,100){\bf (b)}
\includegraphics[width=0.30\linewidth]{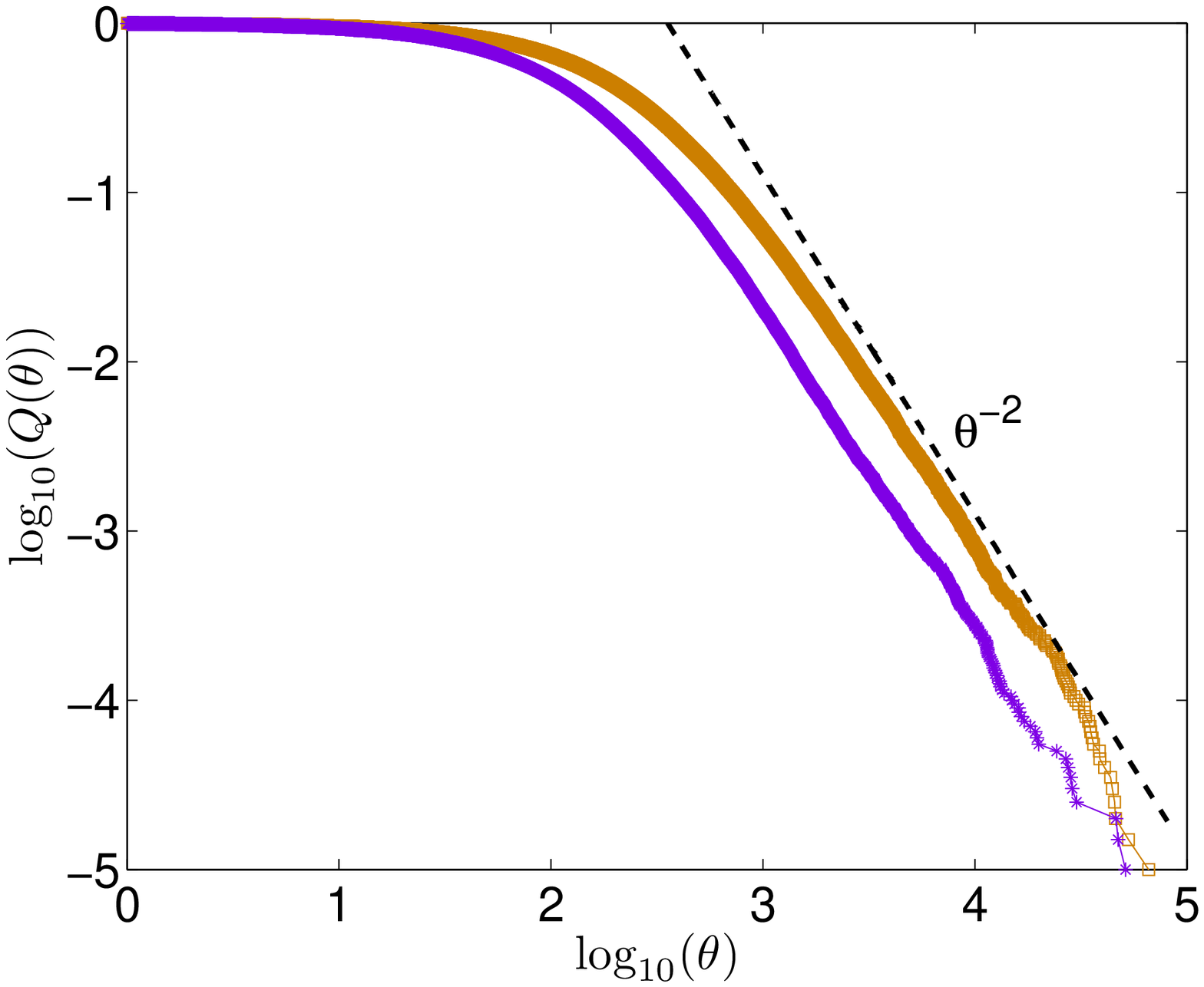}          
\put(-25,100){\bf (c)}
\caption{(Color online) Analogs of (a) Fig.1(b), (b) Fig.2(a), and (c) Fig.2(b), of the main paper,
obtained by using the stochastic model described above (Eqs. (4)-(7)).}
\label{fig:model}
\end{figure*}

\begin{figure*}
\includegraphics[width=0.3\linewidth]{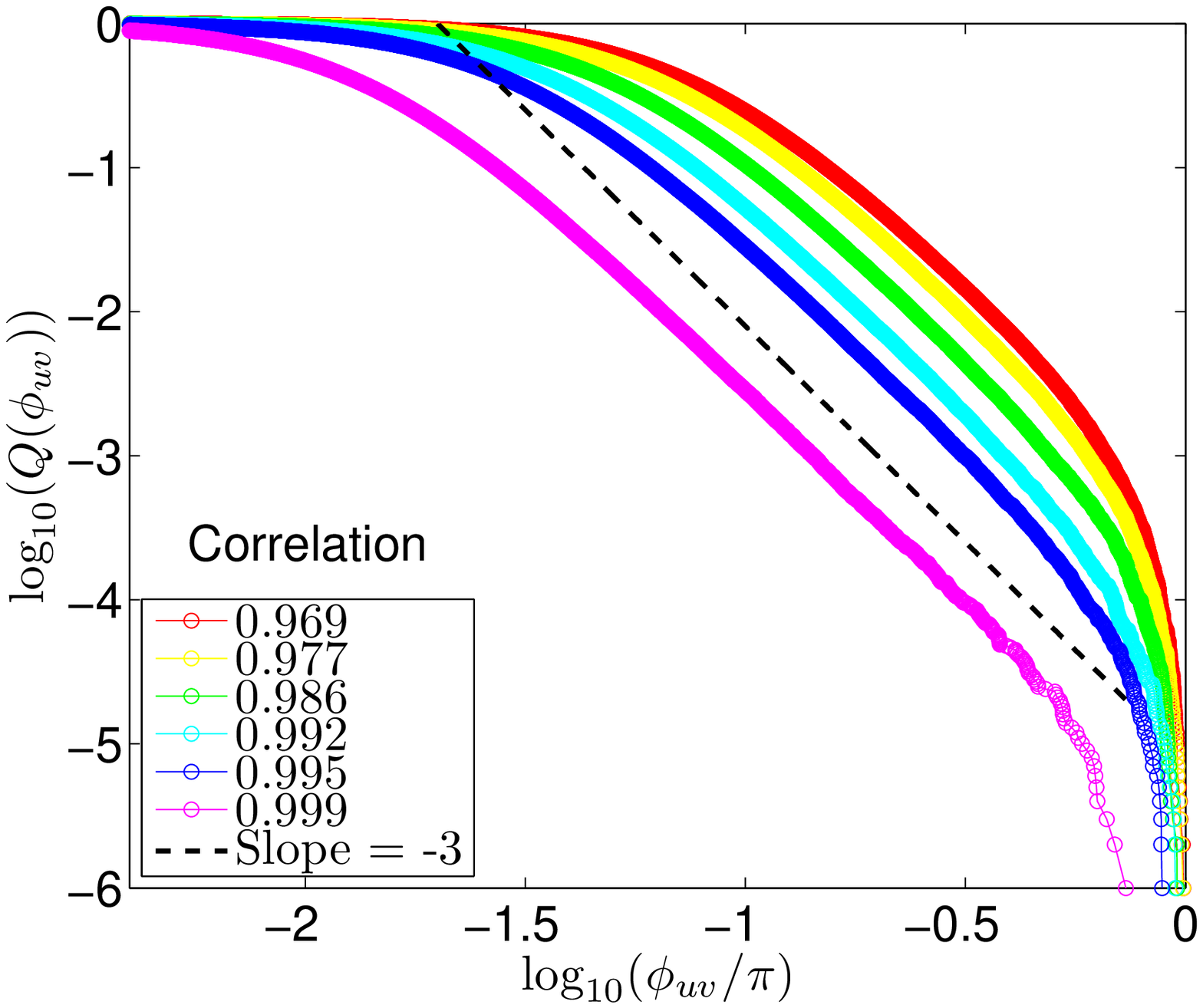}
\put(-25,100){\bf (a)}
\includegraphics[width=0.3\linewidth]{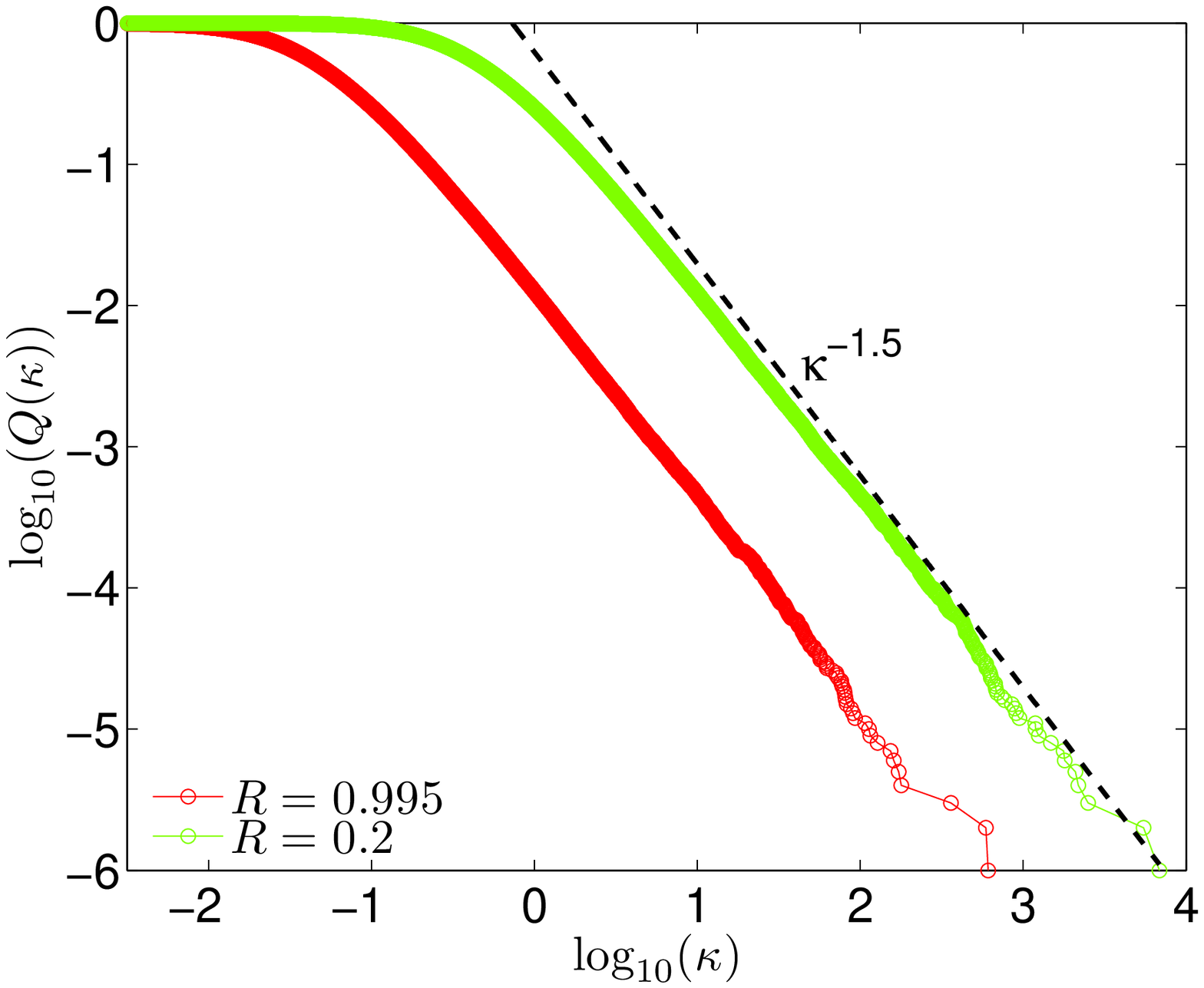}
\put(-25,100){\bf (b)}
\includegraphics[width=0.3\linewidth]{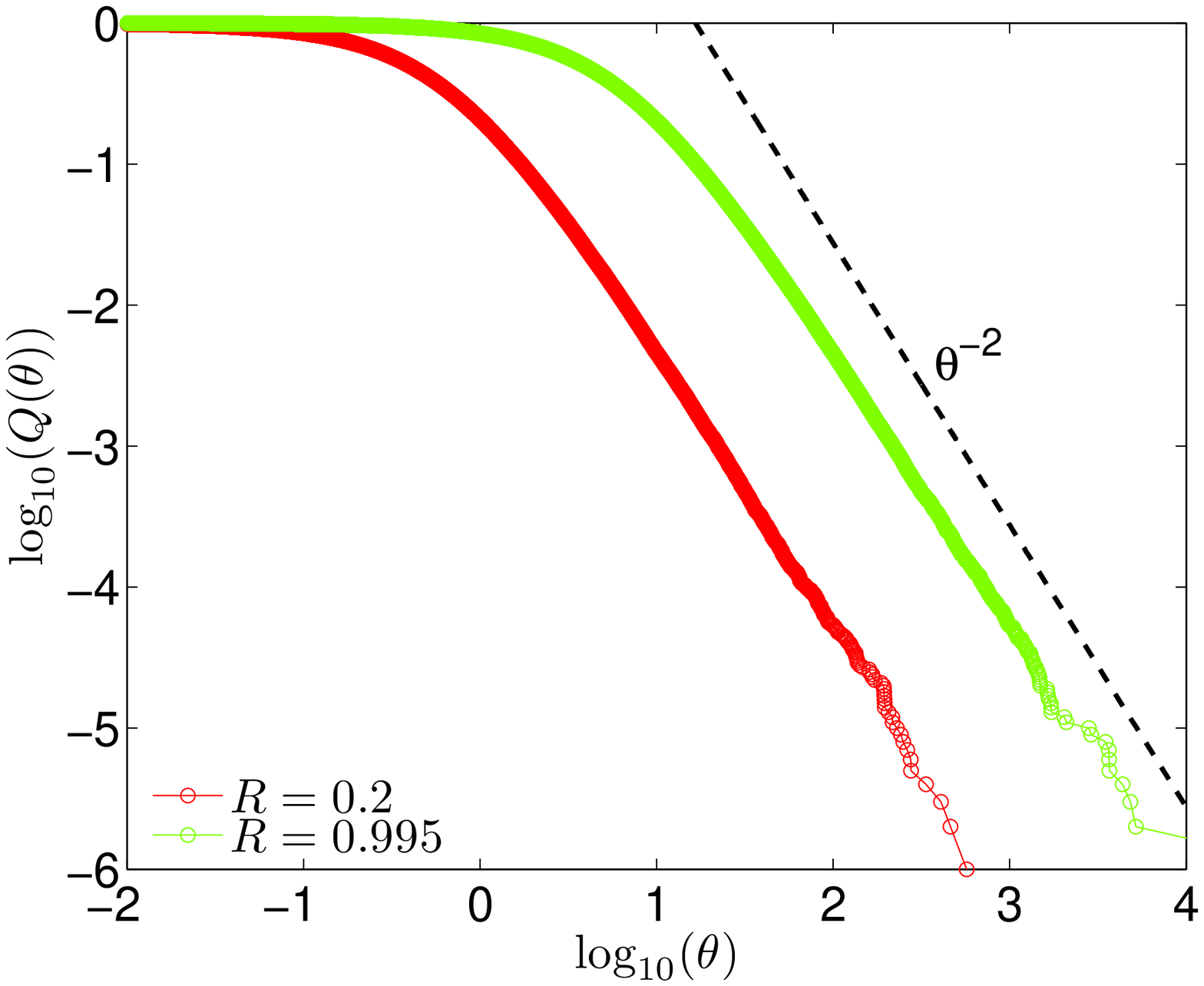}
\put(-25,100){\bf (c)}
\caption{(Color online) Analogs of (a) Fig.1(b), (b) Fig.2(a), and (c) Fig.2(b), of the
main paper, obtained by using the simple models of Eqs. (8)-(9).}
\label{fig:rndm_test}
\end{figure*}

\begin{table}[H]
\caption{Decay constants $\alpha$, $\alpha_t$ and $\alpha_n$ (Eqs.~\ref{eq:acc}), for 
different values of $\St$.}
\centering
\begin{tabular}{c c c c}
\hline
$\St$ & $\alpha$ & $\alpha_t$ & $\alpha_n$ \\
\hline\hline
$0.2$ & $0.31\pm0.08$ &  $0.19\pm0.09$ &  $0.30\pm0.07$ \\
$0.5$ & $0.21\pm0.08$ &  $0.13\pm0.08$ &  $0.20\pm0.09$ \\
$0.7$ & $0.18\pm0.06$ &  $0.11\pm0.09$ &  $0.17\pm0.08$ \\
$1.0$ & $0.15\pm0.04$ &  $0.10\pm0.07$ &  $0.14\pm0.06$ \\
$1.4$ & $0.13\pm0.05$ &  $0.08\pm0.06$ &  $0.11\pm0.06$ \\
\hline
\end{tabular}
\label{table:exponents}
\end{table}

Row (a) of Fig.~\ref{fig:jpdf}
shows joint PDFs of $\theta$ and $a_n$; these joint PDFs demonstrate that large
values of the magnitude of the torsion $\theta$ are associated with small
values of $a_n$.  Row (b) of Fig.~\ref{fig:jpdf} depicts joint PDFs of $\kappa$
and $\theta$. Row (c) of Fig.~\ref{fig:jpdf} shows joint PDFs of $\theta$
and the helicity $H$ of the flow at the position of the particle. These joint
PDFs do not depend strongly on $\St$ and they demonstrate that small and 
the large values of $\theta$ are associated predominantly with $H \simeq H_{rms}$, where
the subscript $rms$ denotes root-mean-square value.

To test the assumption that $\mathcal{P}(a_n,v) \equiv P_\mathrm{an}(a_n) 
P_v(v)$, ~\cite{scagliarini}
we plot the joint PDFs of $a_n$ and $v$ and the product $P_\mathrm{an}(a_n) 
P_v(v)$ side by side, for
different values of $\St$ in Fig.~\ref{fig:jpdfanv}. These joint PDFs show that the
statistical-independence assumption $\mathcal{P}(a_n,v) \equiv P_\mathrm{an}(a_n) 
P_v(v)$, made in ~\cite{geometry3d}, does not hold very well.

{We also use the following stochastic model for the velocity field ${\bf u}$, to obtain all
the statistical properties of particle trajectories that we have discussed in the main
part of this paper. We first define
\begin{eqnarray}
C_{ij}^\pm &=& \cos(i \pm j), \\
S_{ij}^\pm &=& \sin(i \pm j),
\end{eqnarray}
where $i,j \in (x,y,z)$, are the Cartesian co-ordinates. Then we take all three components of
the velocity field $(u_x,u_y,u_z)$ as linear combinations of $C_{ij}^\pm$ and $S_{ij}^\pm$, with
coefficients $\zeta_i$, which evolve in time according to the
following stochastic equation:   
\begin{eqnarray}
\zeta_i(t+\delta t) &=& \zeta_i(t) \exp(-\delta t/T_{cor}) \\
&+& \xi_i(t) \sqrt{\frac{1-\exp(-2 \delta t/T_{cor})}{2}},
\end{eqnarray}
where $T_{cor}$ is the correlation time and $\xi_i(t)$ are chosen from a normal 
distribution~\cite{samriddhi,becoverview}.}
{By using the above model we integrate Eqs.(3) and (4) in the main
paper, to obtain particle trajectories. Figures~\ref{fig:model} (a), (b), and (c) shows the
PDFs of the angle $\phi$, curvature $\kappa$, and torsion $\theta$, respectively, for the
model. We find
that the PDFs $P(\phi)$, $P(\kappa)$ and $P(\theta)$, obtained from the stochastic model
described above, yield the same values for the exponents $\gamma$, $h_\kappa$ and
$h_\theta$ as we obtain from our full DNS in the main paper. However this model does not
yield the form of $n_I(\St)$ (given in Fig. 2 (c) in the main paper) and, therefore, this model
does not yield the exponent $\Delta$.}

Consider now another simple model in which components of the particle and fluid
velocities are \textit{correlated}, 
random Gaussian variates, with
\begin{equation}
\langle u_i \rangle = \langle v_i \rangle = 0; \\
\sigma_{u_i} = \sigma_{v_i} = 1; \\
\label{eq:rndm1}
\end{equation}
here $i= (x,y,z)$, $\langle . \rangle$ represents the average, and $\sigma_{u_i}$ and
$\sigma_{v_i}$ are
the standard deviations of $u_i$ and $v_i$, respectively. We consider the coefficient of correlation
$\rho(u_i,v_j)$ as a function of $\St$, namely,
\begin{equation}
\rho(u_i,v_j) = R(St) \delta_{ij}, 
\label{eq:rndm2}
\end{equation}
here $\rho(u_i,v_j) = \frac{\langle u_i v_j \rangle}{\sigma_{u_i} \sigma_{v_j}}$.
We show numerically that this simple model also gives the same types of $P(\phi)$,
$P(\kappa)$ and $P(\theta)$ as above and values of $\gamma$, $h_\kappa$, and $h_\theta$
that are consistent with our earlier results. To obtain the dependence of these PDFs on
$\St$, we can choose $R(\St)$ in Eq.~\ref{eq:rndm2} to decay with increasing $\St$ as in
Fig. 1(c) in the main paper. 

The simplest stochastic models that we have considered here show that the tails of
$P(\phi)$, $P(\kappa)$, and $P(\theta)$ follow essentially from the correlation
$\rho(u_i,v_j)$. This correlation is dictated by Eqs. (1)-(3) in our DNS in the main paper
or, in the simple stochastic models, by the statistics we use for ${\bf u}$. These simple
stochastic models seem to be adequate for the exponent $\gamma$, $h_\kappa$, and
$h_\theta$ (given our error bars), but not for the exponent $\Delta$.   
 
\vspace{0.5cm}
{\bf Video M1}\\ (\url{https://www.youtube.com/watch?v=lq9X-mdw53o})

This video shows the unit vectors along the directions of the velocity of a heavy inertial 
particle (in red) advected by a turbulent flow, and the velocity of the flow at the position 
of the particle (in  green), and the trajectory of the particle in the neighborhood of the 
particle position (blue dots). This video is from our direct numerical simulation (DNS) of 
the Navier-Stokes equation for the motion of the fluid, and the Stokes-drag equation for the 
motion of the particle. The Stokes number of the particle is one. 

\end{document}